\documentstyle[twocolumn,prb,eqsecnum,aps,floats,epsf]{revtex}

\begin{document}
\draft
\title{Self-Consistent Strong-Coupling-Perturbation Theory for
       \protect\\
       The Anderson Model, Based on Wick's Theorem}
\author{Jan Brinckmann}
\address{Institut f{\"u}r Festk{\"o}rperphysik, Technische Hochschule
         Darmstadt, \protect\\
         Hochschulstr.\ 6, D-64289 Darmstadt, Germany}

\date{  Submitted May 24, 1996 }

\maketitle
\begin{abstract}
%%%****************** Abstract ******************************************
%
A strong-coupling-perturbation theory around the Atomic Limit of the
Anderson model with large $U$ for a localized $f$-orbital coupled to a
conduction-electron band 
is presented. Although an auxiliary-particle representation is {\em
not} used, application of the canonical Wick's theorem is possible and
yields an expansion in the hybridization $V$ via 
dressed skeleton-Feynman diagrams. The Self-Consistent
T-Approximation is constructed as a $\Phi$-derivable
approximation. From a numerical solution of self-consistency
equations the $f$-electron-excitation spectrum is
investigated. Comparison to the Non-Crossing Approximation is made in
virtue of exact formal relations and numerical results. An extension
of this Feynman-diagram approach to the Anderson-lattice model is
indicated, and application within the Local-Approximation scheme
(limit of infinite spatial dimension) is given. 
%
%%%******************* End of abstract **********************************
\end{abstract}
\pacs{ }
%
%
%
%%%********************** body of paper *********************************
%
%%%******************* section starts here ***
\section{Introduction}
\label{sec-intro}
Theories of highly correlated electron systems like Kondo Alloys
\cite{gruzaw78,hewsonbuch}\,, Heavy-Fermion Systems
\cite{fulkelzwi88,greste90} and High-T$_c$ Superconductors
\cite{mmshtsc94} mostly start from model Hamiltonians like the
Anderson or Hubbard model and its extensions 
\cite{and61,hub63,eme87,zharic88}\,. Recently a lot of interest has
also been taken in the multi-channel version of the Kondo model 
\cite{nozbla80}\,, to 
explain non-Fermi-liquid behavior observed in non-magnetic
2-level-impurity systems \cite{ralbur92,raletal94,hetetal94} and
certain U-based 
Heavy-Fermion compounds \cite{cox87,mapetal94}\,. Common to these
models is a single impurity or a lattice of 
impurities endowed with a local interaction. In Hubbard-like
models considered here, electrons on localized $d$- or $f$-orbitals
experience a Coulomb repulsion $U$\,. For the materials mentioned
above $U$ is the largest electronic energy, and conventional 
weak-coupling theories appear not feasible.  
Besides several mainly numerical techniques,
strong-coupling-perturbation theory around the limit of isolated 
localized orbitals (Atomic Limit) is appropriate for
the problem with very large $U$\,. 
The merits of a self-consistent perturbational approach are 
continuous excitation spectra calculated from Green's functions at the real
frequency axis, and the possibility to separate quasi-particle
formation from residual interactions. On the other
hand, (diagram) rules are required which allow for a partial
re-summation of the perturbation series to infinite order. In general
this is difficult to achieve in a strong-coupling approach, where the
Coulomb-interaction operator $\sim U$ is part of the unperturbed
Hamiltonian, whereas the small hybridization between localized orbitals
is taken as the perturbation. 

The earliest strong-coupling-perturbation approach \cite{keikim71}
deals with time-ordered Goldstone
diagrams obtained from an expansion of the Hamiltonian's resolvent
\cite{blodom58} without Wick's theorem and is well
established for the Anderson-impurity model (for a Review see e.g.\ Refs.\
\onlinecite{keimor,bic87,hewsonbuch}). There is nevertheless 
enduring interest in alternative approaches, since an extension to
e.g.\ the Anderson-lattice model imposes difficulties
on the partial re-summation\cite{grekei81,kur85,gre87}\,, owing to the
use of Goldstone diagrams, i.e.\ the loss of Wick's theorem. 
In Slave-Boson theories auxiliary (i.e.\ unphysical) particles are introduced
together with a fluctuating field, which represents the constraint
\cite{note-bic}\,. Conventional many-body technique based on Wick's
theorem can then be used in the calculation of fluctuation corrections
to a saddle point, for impurity and also for lattice models 
\cite{reanew83b,rasdes86,auelev86,millee87,col87,kagyos89,ubblee94a}\,.
However, one has to start from  
the  Mean-Field ground state (reviewed in e.g.\ Refs.\
\onlinecite{newrea87,woel95}), which exhibits broken
gauge symmetry \cite{rea85} with an admixture of unphysical
states which influence sum rules \cite{brigre91,guirag95}\,. 
A variety of alternative diagrammatic approaches have been proposed,
which start directly from the Atomic Limit rather than the Slave-Boson
Mean-Field ground state, like cumulant expansions
\cite{hub66,met91,barcha93,becful88,polbeczev91}\,, 
diagram-techniques for Hubbard operators
\cite{kei68b,hew77,yan89,izylet90,kul93}\,, and 
slave-particle methods where the Mean-Field phase is suppressed in
keeping the constraint exact in each order perturbation theory
\cite{col84,yinkur88}\,.  Other
attempts seek Wick's theorem in the limit of zero temperature
\cite{grojoh89,brigre96}\,. 

The purpose of this paper is to present a self-consistent
skeleton-diagram expansion around the Atomic Limit of the
Anderson-impurity model with large $U$\,, which is built on Wick's
theorem 
in its conventional form \cite{gau60}\,. Neither
auxiliary degrees of freedom nor additional constraints are
introduced. The theory of lattice models is also considered in the light
of this new approach. 

In Sect.\ \ref{sec-wickwick} an expansion \cite{jbepl,jbsces} in the
hybridization $V$ using conventional Feynman diagrams is re-derived
and generalized. The 
Self-Consistent T-Approximation (SCTA) is constructed in Sect.\
\ref{sec-general} as a Conserving Approximation for a set of fermionic
Green's functions, and  is specialized to the Kondo regime in Sect.\
\ref{sec-kondo}\,. Numerical results for the $f$-excitation spectrum are 
presented in Sect.\ \ref{sec-numeric}\,. Here also the relation to the
Non-Crossing Approximation is enlightened via numerical
calculations, exact sum rules and spectral decompositions, 
and a certain limit of the self-consistency (SCTA-) Equations. In
Sect.\ \ref{sec-lattice} we turn to the Anderson-lattice
model and investigate the Feynman-diagram expansion
on the lattice as well as the application of the so-called Local
Approximation. Results are summarized, and conclusions are drawn in Sect.\
\ref{sec-summa}\,. Appendices \ref{app-cano} to \ref{app-sumrules}
contain calculational details not included in the main text.
%
%
%
%%%******************* section starts here *****************************
\section{Effective Hamiltonian and 
         Application of Wick's Theorem}
\label{sec-wickwick}
%%%*******************
%
%
%
%
\subsection{Transformation of The Model Hamiltonian}
\label{sec-trafo}
%%%*******************
%
%
We start from the Anderson Hamiltonian \cite{and61}
\begin{displaymath}
  H = H^c + H^{loc} + H^V_{01} + H^V_{12}
\end{displaymath}
of metallic conduction electrons $H^c$ coupled to a localized $d$- or
$f$-shell $H^{loc}$ through a hybridization term $H^V_{01}+H^V_{12}$\,:
\begin{mathletters}  \label{s2-hamx}
\begin{eqnarray}
  H^c & = &      \label{s2-hband}
    \sum_{k,\sigma}\varepsilon_{k \sigma}
    c^\dagger_{k \sigma} c_{k \sigma}
    \;,\; 
    \\ 
  H^{loc} & = &  \label{s2-local}
      \sum_\sigma \varepsilon^f_\sigma  n^f_\sigma +
      \frac{U}{2}\sum_\sigma n^f_\sigma n^f_{-\sigma}
    \;,\; 
    \\ 
%%%  H^V & = &      \label{s2-decomp}
%%%    H^V_{01} + H^V_{12}
%%%    \;,\; 
%%%    \\ 
  H^V_{01} & = & \label{s2-vpart01}
    \frac{1}{\sqrt{N_{BZ}}} \sum_{k,\sigma}(1-n^f_{-\sigma})
    \left(V_k f^\dagger_\sigma c_{k \sigma} + h.c.\right)
    \;,\; 
    \\ 
  H^V_{12} & = & \label{s2-vpart12}
      \frac{1}{\sqrt{N_{BZ}}} \sum_{k,\sigma}(n^f_{-\sigma})
      \left(V_k f^\dagger_\sigma c_{k \sigma} + h.c.\right)
      \;\,. 
\end{eqnarray}
\end{mathletters}
$n^f_\sigma = f^\dagger_\sigma f_\sigma$\,, 
and $N_{BZ}$ denotes the number of wave vectors $k$ in the 1.\ Brillouin
zone. The one-particle level $\varepsilon^f_\sigma$ of the localized
orbital (designated as 
$f$-level and $f$-orbital, respectively) and conduction-electron
band-structure 
$\varepsilon_{k\sigma}$ for spin $\sigma$ are measured relative to a
fixed chemical potential. 
Conduction electrons form a symmetric band $-D<\varepsilon_k<D$ with cut off
$D$\,, and the $f$-level is situated within the
filled Fermi sea of conduction electrons,
$-D<\varepsilon^f_\sigma<0$\,. Indices $k$ or $\sigma$ will be omitted
in the following, if appropriate. 

The decomposition of the hybridization into $H^V_{01}$ and $H^V_{12}$ is a 
suitable starting point for strong coupling perturbation theory, where
the Atomic $(V_k=0)$ Limit $H^c + H^{loc}$ is taken as the unperturbed
Hamiltonian: The first part $H^V_{01}$ of the perturbation induces
transitions only within the Fock space sector ${\cal H}_{0,1}$ which
consists of all states with empty or singly occupied $f$-orbital. The
second part $H^V_{12}$ mixes ${\cal H}_{0,1}$ and its complement
${\cal H}_{2}$ containing all states with double $f$-occupancy. The
sector ${\cal H}_{2}$ involves the Coulomb repulsion $U$\,, and
according to the hierarchy 
\begin{equation}  \label{s2-hierar}
  V \ll (|\varepsilon^f|, D) \ll U 
\end{equation}
it is convenient to eliminate at first $H^V_{12}$ in leading order $V/U$
through a sequence of two canonical transformations. The calculation is
outlined in Appendix \ref{app-cano} and leads to the Hamiltonian
\begin{equation}  \label{s2-heff}
  H \to H'' = 
    H^c + H^{loc} + H^V_{01} + H^J \;\,.
\end{equation}
In the effective interaction $H^J$ appearing here, 
an exchange integral is simplified for large $U$
and $V_k=V$ to $J_U:=-|V|^2/(\varepsilon^f+U) <0$\,, 
\begin{equation}  \label{s2-exham}
  H^J = 
    \frac{1}{N_{BZ}} \sum_{k,q,\sigma} J_U
    \left( f^\dagger_\sigma c^\dagger_{q,-\sigma}f_{-\sigma}c_{k \sigma}
    + f^\dagger_\sigma c^\dagger_{q,-\sigma}c_{k,-\sigma}f_\sigma 
    \right)
\end{equation}
The spin part of $H^J$ may also be written as 
$-2J_U{\bf s}^c\cdot{\bf s}^f$
and reveals the anti-ferromagnetic coupling of conduction-electron-spin 
density ${\bf s}^c$ to the $f$-orbital spin ${\bf
s}^f$ due to virtual excitations into states with doubly occupied
$f$-orbital. 

In continuing the sequence of canonical transformations we could also
eliminate $H^V_{01}$ and would reproduce the well known Kondo exchange
Hamiltonian involving only spin degrees of freedom with a total
coupling constant 
$J^{tot} = |V|^2/\varepsilon^f + J_U$\,,
equivalent to the result of Ref.\ \onlinecite{schwol66}\,.
We do not pursue this way, which is appropriate if the $f$-level
lies well below the conduction band.

An essential result from the above considerations is the fact that 
the effective Hamiltonian Eq.(\ref{s2-heff}) leaves Hilbert space 
sectors ${\cal H}_{0,1}$ and ${\cal H}_{2}$ decoupled; this enables
the expansion in conventional Feynman diagrams given below.
%
%
%%%*******************
%
\subsection{Wick's Theorem in 
            Strong-Coupling Expansion}
\label{sec-wick}
%%%*******************
%
%
For a study of the Anderson model's one-particle-excitation
spectra and transport properties the quantity of interest is the
$f$-Green's function 
\begin{equation}  \label{s2-fgf1}
    F_\sigma(\tau-\tau') = 
      -\langle {\cal T}\{ f_\sigma(\tau) f^\dagger_\sigma(\tau') \}\rangle
      \;\,.
\end{equation}
The symbol ${\cal T}\{\ldots\}$ orders canonical $f$-operators
$f^{(\dagger)}_\sigma$ according to imaginary time $\tau,\tau'$\,;
thermal expectation values $\langle\ldots\rangle$ and Heisenberg
operators are formed with the Hamiltonian $H''$\,. 

By insertion of 
$1 = (1-P_2) + P_2$\,, $P_2=n^f_\uparrow n^f_\downarrow$
into Eq.(\ref{s2-fgf1})\,, the propagator is split into 
\begin{equation}  \label{s2-fgfsplit}
  F_\sigma(\tau-\tau') = 
      F^{\text{low}}_\sigma(\tau-\tau') + 
      F^{\text{high}}_\sigma(\tau-\tau') \;,\; 
\end{equation}
with a low-energy part $F^{\text{low}}$ and a high-energy part
$F^{\text{high}}$\,,
\begin{eqnarray}
  \lefteqn{
  F^{\text{low}}_\sigma(\tau-\tau') = }       \label{s2-fgf1-low}
    \\  
  & & \nonumber \hspace*{2em} 
      -\langle {\cal T}\{ \left([1-n^f_{-\sigma}]f_\sigma\right)\!(\tau)
                          \left([1-n^f_{-\sigma}]f^\dagger_\sigma\right)\!(\tau')
                          \}\rangle \;,\; 
      \\
  \lefteqn{
  F^{\text{high}}_\sigma(\tau-\tau') = }      \label{s2-fgf1-high}
    \\  
  & & \nonumber \hspace*{2em} 
      -\langle {\cal T}\{ \left(n^f_{-\sigma}f_\sigma\right)\!(\tau)
                          \left(n^f_{-\sigma}f^\dagger_\sigma\right)\!(\tau')
                          \}\rangle 
      \;\,. 
\end{eqnarray}
The projector $P_2$ onto subspace ${\cal H}_2$ commutes with $H''$\,,
\begin{equation}  \label{s2-cons}
  \left[P_2,H''\right] = 0 
    \;,\; 
\end{equation}
i.e.\ double $f$-occupancy is conserved, and therefore 
mixed terms not written in Eq.(\ref{s2-fgfsplit}) vanish.

The low-energy part Eq.(\ref{s2-fgf1-low}) contributes to the
one-particle-excitation spectrum $\rho^f(\omega)$ of $f$-electrons in
the frequency range 
$0\le|\omega|<D$ we are interested in, whereas the spectrum of
$F^{\text{high}}$ is well separated from that by $U$\,, due to excitations into
${\cal H}_2$\,. Thus the high-energy part will be ignored in the
following and we take $F = F^{\text{low}}$\,, that is
\begin{eqnarray}
  \lefteqn{  F_\sigma(\tau-\tau') = }  \label{s2-fgf2}
    \\
  & &  \nonumber \hspace*{1em} 
      -\frac{1}{Z}
      \text{Tr}[ e^{-\beta H''}
      {\cal T}\{\,
      \left(f_{-\sigma}f^\dagger_{-\sigma}f_\sigma\right)\!(\tau)\,
      \left(f_{-\sigma}f^\dagger_{-\sigma}f^\dagger_\sigma\right)\!(\tau')
      \,\} ]
\end{eqnarray}
with the partition function 
$Z = \text{Tr}[ e^{-\beta H''}]$ 
and the inverse temperature $\beta = 1/k_BT$\,. 

Our goal is a perturbation expansion of the $f$-Green's function in
the hybridization $H^V_{01}$ and the exchange coupling term $H^J$ 
appearing in Eq.(\ref{s2-heff})\,. In order to apply Wick's theorem on the
course, two points have to be noted: First, on account of the
conservation law Eq.(\ref{s2-cons}) the Fock-space sector
${\cal H}_2$ involving double $f$-occupancy is completely decoupled
from the dynamics introduced by the perturbation. Second, 
states from ${\cal H}_2$ do not contribute at all to the trace in
Eq.(\ref{s2-fgf2})\,, owing to Eq.(\ref{s2-cons}) and 
$P_2\,(f_{-\sigma}f^\dagger_{-\sigma}f^{(\dagger)}_\sigma) = 
      (f_{-\sigma}f^\dagger_{-\sigma}f^{(\dagger)}_\sigma)\,P_2 = 0$\,.
Thus the interaction term in the local $f$-Hamiltonian $H^{loc}$ is
completely ineffective, and 
we may assign any value to the parameter $U$ in $H^{loc}$\,, or ignore
this interaction term for convenience:
$  H^{loc} \to \widetilde{H}^{loc} = 
    \sum_\sigma \varepsilon^f_\sigma n^f_\sigma \;\,.$
Accordingly we may work with a modified Hamiltonian
$H'' \to \widetilde{H}$\,,
\begin{equation}  \label{s2-htil}
  \widetilde{ H} = \widetilde{ H}^{(0)} + H^V_{01} + H^J
\end{equation}
where the unperturbed part $\widetilde{ H}^{(0)}$ is now bilinear in
band- and $f$-operators,
\begin{equation}  \label{s2-htil0}
  \widetilde{ H}^{(0)} = 
    \sum_{k,\sigma}\varepsilon_{k \sigma}
    c^\dagger_{k \sigma}c_{k \sigma} + 
    \sum_\sigma \varepsilon^f_\sigma f^\dagger_\sigma f_\sigma 
    \;\,.
\end{equation}
Introducing properly normalized expectation values
\begin{equation}  \label{s2-exptil}
  \langle\ldots \widetilde{\rangle} \equiv
    \frac{1}{\widetilde{ Z}}
    \text{Tr}[e^{-\beta \widetilde{ H}}\ldots]
    \;,\;
    \widetilde{ Z} = 
      \text{Tr}[e^{-\beta \widetilde{ H}}]
      \;,\; 
\end{equation}
the one-particle Green's function Eq.(\ref{s2-fgf2}) now reads
\begin{equation}  \label{s2-fgf3}
  F_\sigma(\tau-\tau') = 
    \frac{\widetilde{ Z}}{Z} \widetilde{ F}_\sigma(\tau-\tau') \;,\; 
\end{equation}
with
\begin{eqnarray}
  \lefteqn{ \widetilde{ F}_\sigma(\tau-\tau')  = }  \label{s2-fgftil}
    \\
  & &  \nonumber  \hspace*{2em} 
    - \langle {\cal T}\{
    \left(f_{-\sigma}f^\dagger_{-\sigma}f_\sigma\right)\!(\tau)\,
    \left(f_{-\sigma}f^\dagger_{-\sigma}f^\dagger_\sigma\right)\!(\tau')
    \} \widetilde{\rangle} \;\,.
\end{eqnarray}
In $\widetilde{ F}$ the thermal average as well as time
dependencies of (composite) operators 
$A(\tau) = e^{\tau \widetilde{ H}}A e^{-\tau \widetilde{ H}}$
are determined by the modified Hamiltonian $\widetilde{ H}$\,.
Note that double $f$-occupancy does contribute to $\widetilde{ Z}$ and
therefore $\widetilde{ Z}$ differs from the partition function $Z$\,.

An expansion in connected Feynman diagrams by application of Wick's theorem and
linked-cluster theorem is now apparent, since $\widetilde{F}$ is a
3-particle Green's function of a fermion system, with an unperturbed
Hamiltonian $\widetilde{ H}^{(0)}$ and a  two-particle
interaction $H^V_{01} + H^J$ given through 
Eqs.(\ref{s2-vpart01}) and (\ref{s2-exham})\,. Following e.g.\ Ref.\
\onlinecite{negorl} a diagram of given order in $V$ and
$J_U$ is constructed from bare fermionic Green's functions and the vertices 
shown in Fig. \ref{fig-vertgf}\,: A dashed line represents a local
fermionic Green's function 
\begin{equation}  \label{s2-gtil0}
  \widetilde{ G}^{f(0)}_\sigma(i\omega_l) = 
    \frac{1}{i\omega_l-\varepsilon^f_\sigma}
%%%    \;,\;
%%%    \omega_l = (2l+1)\pi/\beta
    \;,\; 
\end{equation}
with Matsubara frequency $\omega_l=(2l+1)\pi/\beta$\,. 
A full line corresponds to a band electron at the impurity site, with
the $k$-dependence of $V_k$ included,
\begin{equation}  \label{s2-gc0}
  G^c_{\sigma}(i\omega_l) = \frac{1}{N_{BZ}}\sum_k 
    \frac{|V_k|^2}{V^2} \frac{1}{i\omega_l-\varepsilon_{k \sigma}}
    \;\,.
\end{equation}
For counting the order in perturbation theory a real average $V$ 
has been introduced,
\begin{equation}  \label{s2-vauav}
    V^2 := 
      \frac{1}{N_{BZ}}\sum_{k} |V_k|^2 \;\,.
\end{equation}
Vertices originating from the correlated hybridization $H^V_{01}$ are
shown in Fig.\ \ref{fig-vertgf}~(a) and contribute a factor $V$ each,
those corresponding to $H^J$ carry a factor $J_U$\,, and are depicted in  
Fig.\ \ref{fig-vertgf}~(b)\,. All vertices conserve direction of
arrows (i.e.\ particle number), spin $\sigma$ and Matsubara
frequency $\omega_l$\,. Each internal free frequency to be summed up
is accompanied by a $1/\beta=k_BT$ and an unusual convergence factor
$\exp(-i\omega_l 0_+)$\,: its exponent differs in sign from standard
rules, owing to non-normal ordering of $f$-operators
\cite{pap0995-note1} in Eq.(\ref{s2-vpart01})\,. Sums over internal
wave vectors $k$ are 
always independent and already included in the local band-electron
line $G^c$\,. 
The sign $(-1)^{n_c + \chi_{ext} + n_J}$ of a diagram is determined as
usual by the number $n_c$ of closed fermion loops, the ordering
$\chi_{ext}$ of external time labels, and the number $n_J$ of vertices
$\sim J_U$\,. The sign contribution from vertices $\sim V$ is
compensated by the non-normal ordering $(1-n^f)=f f^\dagger$ of
$f$-operators in Eq.(\ref{s2-vpart01})\,. 

An interpretation of vertices and diagrams in physical processes as
indicated in Fig.\ \ref{fig-vertgf} is obvious, but has to be
considered merely as 
illustrative: The dashed line $\widetilde{ G}^{f(0)}$ does {\em not}
represent a true bare $f$-particle $F^{(0)}$ in the unperturbed limit $V=0$\,,
which is given by \cite{hub63}
\begin{displaymath}
  F^{(0)}_\sigma(i\omega_l) = 
    \langle 1-n^f_{-\sigma} \rangle^{(0)} \,
    \widetilde{ G}^{f(0)}_\sigma(i\omega_l)
    \;\,; 
\end{displaymath}
hence the propagator $\widetilde{ G}^{f(0)}$ 
lacks the reduced spectral weight factor of $F^{(0)}$\,. Although this is
only a minor effect here, differences become much more pronounced if 
full propagators are considered. 

For the remainder of this paper we turn to the $U\to\infty$ limit,
that is, vertices $\sim J_U$ in Fig.\ \ref{fig-vertgf}~(b) will be
dropped according to the Hamiltonian 
$\displaystyle
  \widetilde{H} \Rightarrow \widetilde{ H}^{(0)} + H^V_{01}$\,. 
The partition function $Z$ is now given through  
\begin{eqnarray}
  U\to\infty\,:\;\;\; Z & = &  \nonumber 
    \mbox{Tr}[e^{-\beta H''}] \to
    \mbox{Tr}[e^{-\beta \widetilde{ H}}(1-P_2)]
    \\ 
  & = &  \label{s2-zinf}
     \widetilde{ Z} - Z^c\exp(-\beta\sum_\sigma
        \varepsilon^f_\sigma )
     \;\,. 
\end{eqnarray}
Thus $Z$ is accessible via an expansion of $\ln(\widetilde{ Z})$ in
linked diagrams. 
$Z^c$ denotes the partition function of bare conduction electrons. 

The Hamiltonian Eq.(\ref{s2-htil}) in the limit $U\to\infty$\,, i.e.\ $J=0$
is well known \cite{tou70}\,. It has widely
been studied using expansions in the hybridization $H^V_{01}$ via time
ordered (Goldstone) diagrams based on the Resolvent Method
\cite{tou70,note-bic}\,, or equivalent schemes 
involving constraints on auxiliary particles
\cite{col84,yinkur88}\,. Nevertheless 
the possibility to apply the canonical Wick's theorem in a
strong-coupling expansion has not been realized so far. 

Furthermore it is interesting to note that the approach proposed here
is easily transferred to 
the spin-$1/2$ Kondo model \cite{kon64} and 
multi-channel Kondo model \cite{nozbla80}\,, giving
rise to a Feynman-diagram representation of the conduction electron's 
$T$-matrix: The additional states introduced in writing the local spin in
fermions do not contribute to the T-matrix; thus any
`slave-fermion' constraint is unnecessary for a spin-$1/2$ impurity
model, as has already been argued in Ref.\ \onlinecite{abr65}\,. 

%
%
%
%%%******************* section starts here ***
\section{Self-Consistent Approximation: 
         General Aspects}
\label{sec-general}
In the diagrammatic expansion of the 3-particle propagator
$\widetilde{F}$\,,  which covers the dynamics of one-particle
$f$-excitations via 
Eqs.(\ref{s2-fgf3}) and (\ref{s2-fgftil})\,, we encounter the
logarithmic divergences substantial for formation of the Kondo effect
\cite{abr65}\,. This can already be confirmed for the
simplest ladder diagrams shown in Fig.\ \ref{fig-logloops}\,, which
contribute to $\widetilde{ F}$\,. Details concerning
representations of $\widetilde{ F}$ will be given later on. Consider e.g.\
the particle--particle $(pp)$ diagram $\gamma^{pp}(i\nu_k)$\,:
Its analytic contribution at bosonic frequency $i\nu_k=2k\pi/\beta$ is
\begin{displaymath}
  \gamma^{pp}(i\nu_k) \approx
    V^2 \rho^c(0) \int_{-D}^D d \varepsilon\, 
    \frac{ f(\varepsilon) }{i\nu_k-\varepsilon-\varepsilon^f}
    \;,\; 
\end{displaymath}
where $k_BT\ll|\varepsilon^f|$ and $f(\varepsilon+i\nu_k) =
f(\varepsilon)$ denotes the Fermi 
function. Spin degeneracy has been assumed, and a flat 
conduction-band density 
$  \rho^c(\varepsilon) = 
    -\frac{1}{\pi}\mbox{Im}\,G^c(\varepsilon+i0_+) = 
    \frac{1}{2D}\Theta(D-|\varepsilon|) $
inserted. 
After continuation to the real axis, $i\nu_k\to\omega\pm i0_+$\,, the
real part $\mbox{Re}\,\gamma^{pp}(\omega\pm i0_+)$ develops for $T\to
0$ a logarithmic singularity at the $f$-level
$\omega=\varepsilon^f$\,. 
Similar results are obtained for the particle--hole $(ph,
\overline{ph})$ diagrams in Fig.\ \ref{fig-logloops}\,. 

The strategy to be pursued is quite common to known perturbational
strong-coupling approaches\cite{note-bic}\,. Diagrams showing
logarithmic divergences have to be re-summed, whereby a small
Kondo-energy scale in the form 
$T_A = D\exp(\varepsilon^f / 2V^2\rho^c(0))$
is established\cite{rashew84,brakeiliu85}\,. Furthermore
a renormalization of propagators in skeleton
diagrams is necessary\cite{gre83a,kur83,leezha84,col84}\,, which leads
to a smooth 
Abrikosov--Suhl resonance near the Fermi edge in the
$f$-excitation spectrum and turns $T_A$ towards the exact Kondo energy
\cite{kriwilwil80b,tsvwie83,sch89} $T_K$\,. Accordingly we aim at an
approximation for the propagator $\widetilde{ F}$ via
summation of a certain class of dressed skeleton-Feynman diagrams
\cite{lutwar60} to 
infinite order in $V$\,, involving the diagram elements shown in
Fig.\ \ref{fig-logloops}\,. 
%
%%%*******************
%
\subsection{Skeleton-Diagram Expansion}
%%%*******************
%
%
To begin with, consider the construction of proper self
energies. Since the vertices from Fig.\ \ref{fig-vertgf}~(a) present in the
$U\to\infty$ problem allow for connected diagrams with external lines
of arbitrary type, i.e.\ $G^c$ ($c$-line, given in Eq.(\ref{s2-gc0}))
or $\widetilde{ G}^{f(0)}$ ($f$-line, see Eq.(\ref{s2-gtil0})), we get
four self energies designated as  
\begin{equation}  \label{s3-allse}
  \Sigma^{ff}_\sigma(i\omega_l)\,, 
  \Sigma^{fc}_\sigma(i\omega_l)\,, 
  \Sigma^{cf}_\sigma(i\omega_l)\,, 
  \Sigma^{cc}_\sigma(i\omega_l)\,.
\end{equation}
The type of outgoing and
incoming line is indicated in the superscript. As usual, external
lines are removed and irreducible (proper) self-energy diagrams cannot
be split by cutting only one line of {\em any} type. 
Therefore not only a renormalization $\widetilde{ G}^{f(0)}\to
\widetilde{ G}^f$ and $G^c\to \widetilde{ G}^c$ takes place, but also
mixed Green's functions $\widetilde{G}^{fc}$ and $\widetilde{ G}^{cf}$
emerge. The $\widetilde{ G}^{aa'}$ are depicted as the various kinds 
of double lines shown in Fig.\ \ref{fig-matrix}~(a)\,. In the 
imaginary-time domain, these  propagators are defined as
\begin{mathletters}  \label{s3-allgtil}
  \begin{eqnarray}
    \widetilde{ G}^{f}_\sigma(\tau-\tau') & = &  \label{s3-gtilff}
      -\langle {\cal T} \{
      f_\sigma(\tau) f^\dagger_\sigma(\tau') 
                        \} \widetilde{\rangle}
      \;,\; 
      \\ 
    \widetilde{ G}^{fc}_\sigma(\tau-\tau') & = &  \label{s3-gtilfc}
      -\langle {\cal T} \{
      f_\sigma(\tau) C^\dagger_\sigma(\tau') 
                        \} \widetilde{\rangle}
      \;,\; 
      \\ 
    \widetilde{ G}^{cf}_\sigma(\tau-\tau') & = &  \label{s3-gtilcf}
      -\langle {\cal T} \{
      C_\sigma(\tau) f^\dagger_\sigma(\tau') 
                        \} \widetilde{\rangle}
      \;,\; 
      \\ 
    \widetilde{ G}^{c}_\sigma(\tau-\tau') & = &  \label{s3-gtilcc}
      -\langle {\cal T} \{
      C_\sigma(\tau) C^\dagger_\sigma(\tau') 
                        \} \widetilde{\rangle}
      \;,\; 
  \end{eqnarray}
\end{mathletters}
with local operators $f^{(\dagger)}_\sigma$ and 
$  C^{(\dagger)}_\sigma = 
    \frac{1}{\sqrt{N_{BZ}}}\sum_k
    \frac{V^{(\ast)}_k}{V} c^{(\dagger)}_{k \sigma} $\,. 
Green's functions Eq.(\ref{s3-allgtil}) and corresponding self
energies Eq.(\ref{s3-allse}) are now collected in a matrix propagator 
\begin{equation}  \label{s3-matrix-gf}
  \mbox{\bf G}_\sigma(i\omega_l) = 
    \left( \begin{array}{cc}  \displaystyle 
	     \widetilde{ G}^f_\sigma 
               &  \displaystyle
               \widetilde{ G}^{fc}_\sigma 
               \\  \displaystyle
	     \widetilde{ G}^{cf}_\sigma 
               &  \displaystyle
               \widetilde{ G}^{c}_\sigma 
	   \end{array}  \right)(i\omega_l)
\end{equation}
and matrix-self energy
\begin{equation}  \label{s3-matrix-se}
  \bbox{\Sigma}_\sigma(i\omega_l) = 
    \left( \begin{array}{cc}  \displaystyle
             \Sigma^{ff}_\sigma &  \displaystyle
               \Sigma^{fc}_\sigma 
               \\  \displaystyle
             \Sigma^{cf}_\sigma &  \displaystyle
               \Sigma^{cc}_\sigma 
	   \end{array}  \right)(i\omega_l)
  \;\,.
\end{equation}
The bare ($V=0$) matrix $\displaystyle \mbox{\bf G}^{(0)}$ contains
only diagonal elements
$\widetilde{G}^{f(0)}$ and $G^c$ since 
mix-Green's functions $\widetilde{ G}^{fc/cf}$ vanish in the
Atomic Limit. Full and bare matrix-Green's functions are shown in Fig.\
\ref{fig-matrix}~(b)\,, followed by Dyson's equation in Fig.\
\ref{fig-matrix}~(c)\,. 
Diagram rules for matrix propagator and vertex function
as well as the vertex in matrix formulation follow from re-writing the
perturbation $H^V_{01}$ in `spinors' 
$ \Psi_\sigma = 
    \left( f_\sigma \,,\, C_\sigma \right) $\,,
$  \overline{\Psi}_\sigma = 
    \left( f^\dagger_\sigma \,,\, C^\dagger_\sigma \right) $\,.
The resulting vertex is displayed in Fig.\ \ref{fig-matvert}~(a)\,. 
At last, we combine the vertex' two possible
orientations as shown in Fig.\ \ref{fig-matvert}~(b)\,. 
The net result is, that diagram rules for 
unphysical `free energy' $\ln(\widetilde{ Z})$ (see
Eq.(\ref{s2-exptil})), Green's function $\displaystyle
\mbox{\bf G}$\,, self energy $\mbox{\boldmath$\Sigma$}$ and all vertex
functions are quite analogous to the interacting electron gas
\cite{negorl}\,,  with an unusual factor 
\cite{pap0995-note5} $\exp(-i\omega_n 0_+)$ for each internal
frequency $\omega_n$\,. 

Dyson's equation shown in Fig.\ \ref{fig-matrix}~(c) is written 
\begin{displaymath}
  ( \mbox{\bf G}_\sigma(i\omega_l) )^{-1} = 
    ( \mbox{\bf G}^{(0)}_\sigma(i\omega_l) )^{-1} - 
    \bbox{\Sigma}_\sigma(i\omega_l)
\end{displaymath}
and is solved by inversion of a $2\times 2$-matrix. The elements of
$\displaystyle \mbox{\bf G}_\sigma$ obtained this way are noted for later use, 
\begin{mathletters}  \label{s3-dys-all}
  \begin{eqnarray}
    \widetilde{ G}^{f}_\sigma & = &  \label{s3-dys-ff}
      \frac{1}{ [ \widetilde{ G}^{f(0)}_\sigma ]^{-1}
                - \widetilde{ \Sigma}^{f} }
      \\ 
    & &  \text{with}\hspace*{4mm}   \nonumber 
      \widetilde{\Sigma}^{f}_\sigma = 
        \Sigma^{ff}_\sigma + 
        \frac{\Sigma^{fc}_\sigma \, \Sigma^{cf}_\sigma }
             {[ G^c_\sigma ]^{-1} - \Sigma^{cc}_\sigma }
      \\ \nonumber \\
    \widetilde{ G}^{fc}_\sigma  & = &  \label{s3-dys-fc}
      \widetilde{ G}^{f}_\sigma 
      \frac{ \Sigma^{fc}_\sigma }
           { [ G^c_\sigma ]^{-1} - \Sigma^{cc}_\sigma }
      = 
      \frac{ \Sigma^{fc}_\sigma }
           { [ \widetilde{ G}^{f(0)}_\sigma ]^{-1} - \Sigma^{ff}_\sigma }
      \widetilde{ G}^{c}_\sigma 
      \\ \nonumber \\
    \widetilde{ G}^{cf}_\sigma  & = &  \label{s3-dys-cf}
      \frac{ \Sigma^{cf}_\sigma }
           { [ G^c_\sigma ]^{-1} - \Sigma^{cc}_\sigma }
      \widetilde{ G}^{f}_\sigma 
      = 
      \widetilde{ G}^{c}_\sigma 
      \frac{ \Sigma^{cf}_\sigma }
           { [ \widetilde{ G}^{f(0)}_\sigma ]^{-1} - \Sigma^{ff}_\sigma }
      \\ \nonumber \\
    \widetilde{ G}^{c}_\sigma & = &  \label{s3-dys-cc}
      \frac{1}{ [ G^{c}_\sigma ]^{-1}
                - \widetilde{ \Sigma}^{c} }
      \\ 
    & &  \text{with}\hspace*{4mm}   \nonumber 
      \widetilde{\Sigma}^{c}_\sigma = 
        \Sigma^{cc}_\sigma + 
        \frac{\Sigma^{cf}_\sigma \, \Sigma^{fc}_\sigma }
             {[ \widetilde{ G}^{f(0)}_\sigma ]^{-1} - \Sigma^{ff}_\sigma }
  \end{eqnarray}
\end{mathletters}

The 3-particle propagator $\widetilde{ F}_\sigma$ of Eq.(\ref{s2-fgftil})\,,
which is the
quantity that furnishes the $f$-excitation spectrum, is conveniently
expressed through the self energies appearing in Dyson's equation
Eq.(\ref{s3-dys-all})\,: In close analogy to the well known T-matrix
relation \cite{note-bic} exact equations of motion are utilized to
show that 
\begin{equation}  \label{s3-tmat}
  \widetilde{ G}^c_\sigma(i\omega_l) = 
    G^c_\sigma(i\omega_l) + 
    G^c_\sigma(i\omega_l)\, V^2 \widetilde{ F}_\sigma(i\omega_l)\,
    G^c_\sigma(i\omega_l) 
    \;,\; 
\end{equation}
with bare and
renormalized conduction electron Green's function $G^c_\sigma$ and
$\widetilde{ G}^c_\sigma$\,, as defined in Eqs.(\ref{s2-gc0}) and
(\ref{s3-gtilcc}) respectively. In combination with
Eq.(\ref{s3-dys-cc}) we get the exact expression
\begin{equation}  \label{s3-tmat2}
  V^2 \widetilde{ F}_\sigma(i\omega_l) = 
    \frac{ \widetilde{\Sigma}^c_\sigma(i\omega_l) }
         { 1 - G^c_\sigma(i\omega_l) 
               \widetilde{\Sigma}^c_\sigma(i\omega_l) }
    \;\,.
\end{equation}
Alternatively this result is derived from considering the diagram
series of $\widetilde{ F}_\sigma$ directly. 
Apparently $V^2 \widetilde{ F}$ equals the {\em im}\/proper self
energy of $\widetilde{ G}^c$\,. Since $\widetilde{ F}$ stands for a
3-particle propagator, it contains diagrams reducible with respect
to band lines $G^c$ (i.e.\ which can be split by cutting a single
$G^c$) compliant to particle conservation, in contrary to any two-particle
Green's function. 
%
%%%*******************
%
\subsection{Mean-Field Theory and 
            Infinite Ladder Summation}
%%%*******************
%
%
The formalism developed so far serves as the basis for defining 
self-consistent approximations: An expansion of the
matrix-self energy $\bbox{\Sigma}$ in skeleton diagrams with 
matrix-type lines and vertices leads 
to {\em consistent} expressions for its elements $\Sigma^{a'a}$\,,
$a'a=ff,cf,fc,cc$\,. These fully determine the self-consistency problem
via Dyson's equation Eq.(\ref{s3-dys-all}) as well as the one-particle
$f$-Green's function via Eqs.\ (\ref{s3-tmat2}) and (\ref{s2-fgf3})\,. 
As a starting point we take the exact representation \cite{xray2,bicbuch}
for $\bbox{\Sigma}$ given in Fig.\ \ref{fig-selfbethe}\,. To
prevent over-counting the lowest order contribution (`Hartree
diagram') is written explicitly, whereas all higher orders are
absorbed into the irreducible vertex function \cite{note-negorl}
$\Gamma^2_P$\,. 

At mean-field level only the `Hartree diagram' in Fig.\ \ref{fig-selfbethe}
is taken into account, which leads to a self energy
$\bbox{\Sigma}$  independent of frequency $\omega_l$\,. The
self-consistency equations resulting from Dyson's equation (\ref{s3-dys-all})
can be solved analytically \cite{pap0995-note7} at $T\to 0$\,. They possess,
however, only a trivial solution:
\begin{eqnarray*}
  \Sigma^{ff}_\sigma(i\omega_l) & = & 
    2V\langle f^\dagger_{-\sigma}C_{-\sigma} \widetilde{\rangle} 
    \hspace*{4mm} \to 0 \;,\; 
    \\ 
  \Sigma^{cf}_\sigma(i\omega_l) & = & 
    \Sigma^{fc}_\sigma(i\omega_l) = 
    V\langle 1-n^f_{-\sigma} \widetilde{\rangle} 
    \hspace*{4mm} \to 0 \;,\; 
    \\ 
  \Sigma^{cc}_\sigma(i\omega_l) & = & 0
    \;\,.
\end{eqnarray*}
That is, the unperturbed ($V=0$) Atomic Limit is recovered in
mean-field theory. This result fits into the discussion of the Kondo
regime given in Sect.\ \ref{sec-kondo} below: The lowest order
diagram (Fig.\ \ref{fig-selfbethe} left) considered here belongs to a
whole class of skeleton diagrams which are negligible in the Kondo regime. 

In going to higher orders, attention has
to be payed to symmetry relations 
fulfilled by the vertex function $\Gamma^2_P$\,: In Fig.\
\ref{fig-selfbethe} the self energy $\bbox{\Sigma}$ has been given 
in a way that $\Gamma^2_P$ enters $\bbox{\Sigma}$ in the
particle--particle $(pp)$ representation 
$\Gamma^2_P(1',2';1,2)$\,, with arbitrary labels attached to incoming
$(1,2)$ and outgoing $(1',2')$ lines. 
Correspondingly it can be analyzed in the $(pp)$-channel, i.e.\
re-written through the Bethe-Salpeter equation 
\begin{eqnarray}
  \lefteqn{ \Gamma^2_P(1',2';1,2) =
    \Gamma^{pp}(1',2';1,2) }    \label{s3-bethe-pp}
    \\
  & & \;\;\;  \nonumber
    \mbox{}+ \Gamma^{pp}(1',2';\overline{1},\overline{2})
    \,G(\overline{1},\overline{1}')
      G(\overline{2},\overline{2}')\,
    \Gamma^2_P(\overline{1}',\overline{2}';1,2)
    \;,\; 
\end{eqnarray}
involving a kernel $\Gamma^{pp}$ which is two-particle irreducible
only in the $(pp)$-channel. 
An integration 
$\int d\overline{1}\,d\overline{2}\,d\overline{1}'\,d\overline{2}'$
is implicit, where e.g.\ $\int d\overline{1}$ is shorthand for
\mbox{$\int_0^\beta d\overline{\tau}_1$}\,
\mbox{$\sum(\overline{\sigma}_1=\pm 1)$}\,
\mbox{$\sum(\overline{a}_1=f,c)$}\,. 
However, the self energy could also be written with reversed arrows on
the spin $(-\sigma)$ loop in Fig.\ \ref{fig-selfbethe}\,, so that 
$\Gamma^2_P$ would be replaced by its particle--hole $(ph)$
representation $\Gamma^2(1',1;2,2')$\,. Now two alternative
Bethe--Salpeter equations are apparent, where $\Gamma^2$ is analyzed in the 
 $(ph)$- or $(\overline{ph})$-channel with respective kernels
$\Gamma^{ph}$ and $\overline{\Gamma}^{ph}$\,,  
\begin{mathletters}  \label{s3-bethe-phall}
\begin{eqnarray}
  \lefteqn{ \Gamma^2(1',1;2,2')= 
    \Gamma^{ph}(1',1;2,2') }  \label{s3-bethe-ph}
    \\
  & & \;\;\;  \nonumber
    \mbox{}+ \Gamma^{ph}(1',1;\overline{2},\overline{2}')
    \,G(\overline{2},\overline{1}')
      G(\overline{1},\overline{2}')\,
    \Gamma^2(\overline{1}',\overline{1};2,2')
    \;\,,\;\,
    \\ \nonumber \\
  \lefteqn{ \Gamma^2(1',1;2,2')= 
    \overline{\Gamma}^{ph}(1',1;2,2') }   \label{s3-bethe-phq}
    \\
  & & \;\;\;  \nonumber
    \mbox{}+ \overline{\Gamma}^{ph}(1',\overline{1};2,\overline{2}')
    \,G(\overline{1},\overline{1}')
      G(\overline{2},\overline{2}')\,
    \Gamma^2(\overline{1}',1;\overline{2},2')
    \;\,.
\end{eqnarray}
\end{mathletters}
Additionally an interchange of the two incoming lines of $\Gamma^2_P$ in Fig.\
\ref{fig-selfbethe} or, if the particle-hole representation has been
chosen, of $\Gamma^2$ does not alter \mbox{\boldmath$\Sigma$}\,, in
virtue of the corresponding anti-symmetry of $\Gamma^2_P$\,,
$\Gamma^2$\,. In combination with the strict equality of
particle--particle and particle--hole representations this
anti-symmetry is expressed through the so-called Crossing Relations
\cite{bicbuch} 
\begin{eqnarray} 
  \lefteqn{
  \Gamma^2_P(1',2';1,2) = 
    -\Gamma^2_P(1',2';2,1) }   \label{s3-cross}
    \\  
  & & \hspace*{12mm}  \nonumber
    = \Gamma^2(1',1;2,2') = 
    -\Gamma^2(1',2;1,2') \;\,.
\end{eqnarray}
These should be fulfilled by any approximation. 

A ladder-type approximation to $\Gamma^2_{(P)}$ may be obtained from Eqs.\
(\ref{s3-bethe-pp}) and (\ref{s3-bethe-phall}) with the kernels
$\Gamma^{pp}$\,, $\Gamma^{ph}$\,, $\overline{\Gamma}^{ph}$ each replaced by
the bare vertex shown in Fig.\ \ref{fig-matvert}~(b)\,. An
iteration of the Bethe--Salpeter equations then yields three different
ladder approximations for the vertex function, with diagrams from only
one of the three inequivalent channels included. Taking solely one of them
into account, with reference e.g.\ to RPA, does certainly not comply
to Crossing Relations. 
It is more suitable to add all three ladder sums and to remove the
lowest order contribution (i.e.\ the bare vertex) twice afterwards,
since it has been counted three times. 
The resulting vertex function $\Gamma^2_P = \Gamma^2$ fulfills Crossing 
Relations (\ref{s3-cross}) and contains no over-counted diagram. 
Furthermore it includes all types of loops from Fig.\
\ref{fig-logloops} to infinite order. 
Eventually, insertion into Fig.\ \ref{fig-selfbethe} leads to the
skeleton approximation for the self energy $\bbox{\Sigma}$ displayed in Fig.\
\ref{fig-selftma}\,. 

The ladder summation obtained is considerably simple, since solely
bosonic propagators representing ladder sums have to be treated
self-consistently. A more systematic approach based on
Parquet equations \cite{xray2,bicbuch,jbdiss} 
would involve `true' retarded vertex functions. 
Additionally the approximation is $\Phi$-derivable, i.e.\ a functional
$\widetilde{\Phi}$ can be found \cite{jbdiss} with the property
\cite{bay62} 
\begin{equation}  \label{s3-phi}
  \left( \bbox{\Sigma}_\sigma(\tau'-\tau) \right)^{a'a} = 
    \frac{\delta\,\widetilde{\Phi}[ \mbox{\bf G} ] }
         {\delta\,\left( \mbox{\bf G}_\sigma(\tau-\tau')\right)^{a a'}}
    \;\,.
\end{equation}
Thus {\em all} self energies $\Sigma^{a'a}$\,, $a'a=ff,cf,fc,cc$ can be
obtained by variation of {\em one} functional
$\widetilde{\Phi}$\,. As a consequence the ladder approximation is
thermodynamically consistent \cite{bay62}\,, i.e.\ expectation values
$\langle\ldots \widetilde{\rangle}$ with a {\em tilde} are safely
computed from corresponding Green's functions. 
Since the diagrammatic representation of $\widetilde{\Phi}$ resembles
that of the T-matrix theory of the electron gas \cite{lutwar60} the
self energy Fig.\ \ref{fig-selftma} will be referred to as the
Self-Consistent T-Approximation (SCTA). 
%
%
%
%%%******************* section starts here ***
\section{Self-Consistent Scheme and 
         Excitation Spectrum in The Kondo Regime}
\label{sec-kondo}
So far an approximation for the Anderson model in the strongly correlated
limit $U\to\infty$ has been derived, which does not rely on any 
specific parameter range such as the Kondo or mixed-valence
regime. In the following the Kondo regime will be considered further,
where the coupled self-consistency equations given through
the matrix \mbox{\boldmath$\Sigma$} and  Dyson's equation
Eq.(\ref{s3-dys-all}) may be simplified using a non-perturbative small
parameter.
%
%%%*******************
%
\subsection{Small Parameter}
\label{sec-small}
%
%%%*******************
%
%
The small parameter available in the Kondo regime is given by the
c-number $Z/\widetilde{ Z}$\,, which relates physical quantities like
the one-particle $f$-Green's function $F$ to
its counterpart indexed by a {\em tilde}\/, as introduced in 
Eq.(\ref{s2-fgf3})\,: 
\begin{equation}  \label{s4-fgf3x}
  F_\sigma(\tau-\tau') = 
    \frac{\widetilde{ Z}}{Z}
    \widetilde{ F}_\sigma(\tau-\tau') \;\,.
\end{equation}
At $U\to\infty$ the prefactor is given through Eq.(\ref{s2-zinf})\,,
\begin{equation}  \label{s4-zinfx}
  \widetilde{ Z}/Z = 
    1 + (Z^c/Z)\exp(-2\beta \varepsilon^f) \;\,.
\end{equation}
We estimate the order of magnitude of the partition function $Z$ via 
\begin{displaymath}
  Z/Z^c = 
    \exp(-\beta(\Omega-\Omega^c)) 
    \approx \exp(-\beta( E_G - E^c )) 
    \;,\; 
\end{displaymath}
where the free energy $\Omega$ has been replaced at low temperature
$T\to 0$ by the ground-state
energy $E_G$\,. Quantities $Z^c, \Omega^c, E^c$ correspond to bare
conduction electrons. As is well known,  
hybridization lowers the Atomic Limit's $(V=0)$ ground-state
energy $E^{(0)}_G = \varepsilon^f + E^c$ only by the small amount of
the Kondo energy $T_K$ to $E_G \approx E^{(0)}_G - T_K$\,. Thus the
estimate $Z/Z^c\approx\exp(-\beta(\varepsilon^f-T_K))$ holds, leading
to 
\begin{equation}  \label{s4-kappa}
  \frac{\widetilde{ Z}}{Z} 
    = 1 + \frac{1}{\kappa}
    \approx \frac{1}{\kappa}
    \;,\;
    \kappa := \exp(\beta(\varepsilon^f+T_K)) 
    \;\,.  
\end{equation}
This provides a non-perturbative parameter $\kappa\ll 1$\,,
which is exponentially small in the Kondo regime 
$\varepsilon^f<0$\,, $T_K\ll|\varepsilon^f|$\,,
$k_BT\ll|\varepsilon^f|$\,. 
From Eq.(\ref{s4-fgf3x}) it now follows that $\widetilde{ F}$ is of very
small magnitude $\sim(\kappa)^1$ (or smaller), 
with $F$ being an intensive thermal expectation value $\sim(\kappa)^0=1$
(or smaller)\,. 

In a similar way information is gained on the mix-Green's functions
$\widetilde{ G}^{cf/fc}$\,: 
Following the arguments given after Eq.(\ref{s2-fgf2})\,, the relation 
\begin{displaymath}
  \langle {\cal T}\{ C_\sigma(\tau) f^\dagger_\sigma(\tau') \}\rangle =
    \frac{\widetilde{ Z}}{Z}
    \langle {\cal T}\{ C_\sigma(\tau) f^\dagger_\sigma(\tau') \}
      \widetilde{\rangle}
\end{displaymath}
holds, together with its complex conjugate involving $f_\sigma$ and
$C^\dagger_\sigma$ defined below Eq.(\ref{s3-allgtil})\,. 
Thus the off-diagonal elements of $\mbox{\bf G}$ in 
Eq.(\ref{s3-matrix-gf}) are of order $\sim(\kappa)^1$\,,
\begin{equation}  \label{s4-mixsmall}
  \widetilde{ G}^{fc}_\sigma(\tau-\tau') \sim (\kappa)^1
    \;,\;
  \widetilde{ G}^{cf}_\sigma(\tau-\tau') \sim (\kappa)^1
    \;\,. 
\end{equation}
It is also useful to consider directly the spectral weight of $F_\sigma$\,,
$\displaystyle   \langle 1-n^f_\sigma \rangle = 
    (\widetilde{ Z}/Z)
    \langle 1-n^f_\sigma \widetilde{\rangle}$\,. 
It follows that
\begin{equation}  \label{s4-ffsmall}
  \langle 1-n^f_\sigma \widetilde{\rangle} = 
    \int_{-\infty}^\infty d\omega\,
    \widetilde{\rho}^f_\sigma(\omega)
    [1 - f(\omega)]
    \sim \kappa
    \;,\; 
\end{equation}
with the Fermi function $f(\omega)$ and the spectral function
\begin{equation}  \label{s4-rhotildef}
  \widetilde{\rho}^f_\sigma(\omega) = 
    -\case{1}{\pi}\mbox{Im}\,\widetilde{ G}^f_\sigma(\omega+i0_+)
   >0 
   \;\,. 
\end{equation}
Since the integrand in Eq.(\ref{s4-ffsmall}) is positive, the function
\begin{equation}  \label{s4-rhosmall}
    \widetilde{\rho}^f_\sigma(\omega)
    [1 - f(\omega)]
    \sim (\kappa)^1 
\end{equation}
may serve as a small parameter, too. This property can also be derived  
from the fact that the spectrum $\widetilde{\rho}^f$ develops a
threshold behavior  at energy $E^{Th}\approx\varepsilon^f+T_K$ when
the temperature goes to zero: In Appendix \ref{app-sumrules} it is
shown that 
$\widetilde{\rho}^f(\omega) \sim \Theta(E^{Th}-\omega)$
at $T=0$\,. It is expected that $\widetilde{\rho}^f(\omega)$ 
shows a singularity as $\omega$ approaches the threshold
$E^{Th}$ from below. Nevertheless, at finite temperature the
singularity is removed, and the left hand side of
Eq.(\ref{s4-rhosmall}) shows its maximum at $\omega\approx E^{Th}$
with a value 
$\widetilde{\rho}^f(E^{Th})\exp(\beta E^{Th}) \approx \kappa \ll 1$\,.

With the small quantities Eqs.(\ref{s4-mixsmall}) and (\ref{s4-rhosmall})
at hand, self-energy diagrams of leading order in
$\kappa$ can be separated from those of higher order in $\kappa$\,, which
will be neglected in the Kondo regime. For that purpose we return to the
explicit notation using the vertices and Green's functions shown in
Figs.\ \ref{fig-vertgf}~(a)\,, \ref{fig-matrix}~(a)\,, and dissolve each
matrix-type diagram from Fig.\ \ref{fig-selftma} into its
contributions to $\Sigma^{a'a}\,, a'a=ff,fc,cf,cc$\,. 
Whereas mix-Green's functions 
contribute directly via Eq.(\ref{s4-mixsmall}) a factor $\kappa$ each, 
Eq.(\ref{s4-rhosmall}) is used within a {\em loop theorem} proven in
Appendix \ref{app-loop}\,: A {\em closed path} in a skeleton diagram to
$\Sigma^{a'a}$\,, which runs exclusively on equally directed 
Green's functions $\widetilde{ G}^f$ (double dashed
lines), is sufficient
for a prefactor $\kappa$ to that diagram. Vertices 
may be passed in arbitrary fashion from any incoming
dashed line to any outgoing dashed line. Accordingly a {\em closed path} in
general involves lines of either spin direction. As an example two
diagrams are shown in Fig.\ \ref{fig-omitted} (a) on the top and
bottom, which possess one and two non-overlapping 
closed paths respectively. By application of the {\em loop theorem}
these are of order $\sim(\kappa)^1$ and $\sim(\kappa)^2$\,. 
Closed paths are indicated by dotted lines; in the diagram on the top only
one of two equivalent paths is marked as an arbitrary choice. 

The resulting subset of ladder diagrams to be considered in the Kondo regime
is displayed in Fig.\ \ref{fig-kondself} for the diagonal elements of
\mbox{\boldmath$\Sigma$}\,, i.e.\ $\Sigma^{ff}$ and
$\Sigma^{cc}$\,. 
$\Sigma^{ff}$ is $\sim(\kappa)^0=1$ because no closed path as 
described above can be found in the diagrams shown in Fig.\
\ref{fig-kondself} (top), nor are any mix-Green's functions $\widetilde{
G}^{fc}$ or $\widetilde{ G}^{cf}$ present. 
The diagrams of $(pp)$-type not shown here as well as all $(ph)$- and
$(\overline{ph})$-diagrams to $\Sigma^{ff}$ are at least 
$\sim(\kappa)^1$ by application of the {\em
loop theorem}\,, i.e.\ vanishing small in the Kondo regime. Fig.\
\ref{fig-omitted} (a) displays examples. Also any diagram containing
at least one mix-Green's function $\widetilde{ G}^{fc}$ or
$\widetilde{ G}^{cf}$  is $\sim(\kappa)^1$ or smaller by
Eq.(\ref{s4-mixsmall}) and can be neglected. Thereby all diagrams of odd
order $V^{2n+1}$ are unimportant, as illustrated in Fig.\
\ref{fig-omitted} (b)\,. 
$\Sigma^{cc}$ on the other hand is in leading order $\sim(\kappa)^1$\,: 
Every diagram of the $(pp)$-type ladder sum indicated in Fig.\
\ref{fig-kondself} (bottom) shows one {\em closed path} on equally directed
dashed lines. All $(ph)$- or $(\overline{ph})$-type
ladder graphs to $\Sigma^{cc}$ not shown here contain at least {\em two }
non-overlapping {\em closed paths\/}. These contribute at order
$\sim(\kappa)^2$ 
and are omitted, as well as all diagrams containing at least
{\em two} mix-Green's functions $\widetilde{G}^{cf/fc}$\,. Furthermore,
it can be seen that all remaining ladder 
graphs to $\Sigma^{cc}$ with only {\em one}
mix-Green's function contain in addition at least one {\em closed path} on
dashed lines and thus are also negligible. 

The above-mentioned simplification of the self-consistency problem
posed through Dyson's 
equations Eq.(\ref{s3-dys-all}) follows from the fact that $\Sigma^{ff}$ is
$\sim 1$ in leading order $\kappa$\,, whereas $\Sigma^{cc}$ and
$f$--$c$-mixing self energies are small, 
\begin{equation}  \label{s4-orders}
  \Sigma^{ff} \sim 1 \;,\;
  \Sigma^{cf} \sim 
  \Sigma^{fc} \sim(\kappa)^1 \;,\;
  \Sigma^{cc} \sim(\kappa)^1 \;\,.
\end{equation}
This is shown in Appendix \ref{app-loop} to hold in general for the
Kondo regime. Eq.(\ref{s3-dys-all}) now reduces to
\begin{equation}   \label{s4-skff}
    \widetilde{ G}^f_\sigma = 
      \left(\left[ \widetilde{ G}^{f(0)}_\sigma \right]^{-1}
          - \Sigma^{ff}_\sigma \right)^{-1}
    \sim 1
\end{equation}
and 
$\displaystyle \widetilde{ G}^c_\sigma=G^c_\sigma\sim 1$\,, 
$\displaystyle \widetilde{ G}^{fc}_\sigma\sim \widetilde{
   G}^{cf}_\sigma\sim(\kappa)^1$\,. 
Therefore $\widetilde{G}^f$ is renormalized self-consistently through
$\Sigma^{ff}$ given in Fig.\ \ref{fig-kondself}\,, whereas the
conduction-electron Green's function $G^c$ remains bare. 
The magnitude of mix-Green's functions $\widetilde{ G}^{fc/cf}$ 
is reproduced consistently. The latter need not be considered in the Kondo
regime. 

The local $f$-Green's function follows from Eqs.(\ref{s4-fgf3x}) and
(\ref{s3-tmat2}), (\ref{s3-dys-cc}) with (\ref{s4-orders}) as 
\begin{equation}  \label{s4-fgf4}
  F_\sigma(i\omega_l) = 
    \frac{\widetilde{ Z}}{Z}
    \frac{\Sigma^{cc}_\sigma(i\omega_l)}{V^2}
    \;\,.
\end{equation}
Although $\Sigma^{cc}\sim(\kappa)^1$ is unimportant for
the renormalization of unphysical propagators, it determines the
physical $f$-Green's function $F$\,, which 
in combination with the prefactor Eq.(\ref{s4-kappa}) is of order $1$.
The property of \mbox{\boldmath$\Sigma$} being $\Phi$-derivable stated  
at the end of the last section, together with Eq.(\ref{s4-fgf4}) takes
the form  
\begin{displaymath}
  \Sigma^{ff}_\sigma(i\omega_l) = 
    \frac{\partial \widetilde{\Phi}[\widetilde{ G}^f, G^c] }
         {\partial \widetilde{ G}^f_\sigma(i\omega_l) }
    \;,\;
  F_\sigma(i\omega_l) = 
    \frac{\widetilde{ Z}}{Z}
    \frac{\partial \widetilde{\Phi}[\widetilde{ G}^f, G^c] }
         { V^2\,\partial G^c_\sigma(i\omega_l) }
    \;\,. 
\end{displaymath}
This corresponds to analogous expressions
derived within Resolvent-Perturbation Theory \cite{kur83}\,. 
%
%%%*******************
%
\subsection{Self-Consistency Equations}
%
%%%*******************
%
%
In order to derive analytical expression for self energies, two
`ladder elements' are defined as shown in Fig.\
\ref{fig-ladders}\,: $\widetilde{\pi}(i\nu_k)$ can be understood as an
effective bosonic propagator with external frequency
$\nu_k=2k\pi/\beta$\,, which is further renormalized via a `self
energy' $\widetilde{ \sigma}(i\nu_k)$ to 
\begin{equation}  \label{s4-pifull}
  \widetilde{\Pi}(i\nu_k) = 
    \frac{\widetilde{\pi}(i\nu_k)}
         {1 - \widetilde{\pi}(i\nu_k)\,\widetilde{ \sigma}(i\nu_k) }
    \;\,. 
\end{equation}
Spin degeneracy is assumed here and in the following, although
$\widetilde{\Pi}$ is spin-independent in general. 
The formula corresponding to Fig.\ \ref{fig-kondself} (top) then reads 
\begin{equation}  \label{s4-sigff}
  \Sigma^{ff}(i\omega_l) = 
    \case{-1}{\beta}\sum_n
    V^2 G^c(i\omega_n) \widetilde{\Pi}(i\omega_n + i\omega_l)
\end{equation}
with external and free fermionic frequencies $\omega_l$ and
$\omega_n=(2n+1)\pi/\beta$ respectively. An overall sign results from
the single closed fermion loop in every diagram. With the
local DOS of conduction electrons (see Eqs.(\ref{s2-gc0}) and
(\ref{s2-vauav}))\,, 
\begin{equation}  \label{s4-cdos}
  \rho^c(\varepsilon) = 
    \frac{1}{N_{BZ}} \sum_k
    \frac{|V_k|^2}{V^2} \delta(\varepsilon-\varepsilon_k)
    \;,\; 
\end{equation}
and a spectral function of the effective propagator $\widetilde{\Pi}$\,,
\begin{equation}  \label{s4-rhopi}
  \widetilde{\rho}^\Pi(\varepsilon) = 
    -\case{1}{\pi} \mbox{Im}\, \widetilde{\Pi}(\varepsilon+i0_+)
    \;,\; 
\end{equation}
the frequency sum in Eq.(\ref{s4-sigff}) is performed through a
contour integration, leading to 
\begin{eqnarray}
  \lefteqn{ \Sigma^{ff}(i\omega_l) = }   \label{s4-sigff2}
    \\ \nonumber 
  & & 
    V^2
    \int\!\!\!\int d \varepsilon \, d \varepsilon' \,
    \rho^c(\varepsilon) \widetilde{\rho}^\Pi(\varepsilon') 
    \frac{ [1-f(\varepsilon)] + [-1-g(\varepsilon')]}
         {i\omega_l + \varepsilon - \varepsilon' }
    \;\,.
\end{eqnarray}
A Bose function $g(\varepsilon')$ appears through
$f(\varepsilon'-i\omega_l) = -g(\varepsilon')$\,.
In the numerator a $1-1=0$ has been added, in order to utilize a
property similar to Eq.(\ref{s4-rhosmall})\,, 
\begin{equation}  \label{s4-pismall}
  \widetilde{\rho}^\Pi(\varepsilon)
  [-1-g(\varepsilon)] \sim (\kappa)^1
  \;\,.  
\end{equation}
Thereby the second term in Eq.(\ref{s4-sigff2}) is of 
order $\sim(\kappa)^1$ and will be neglected in favor of the first
term $\sim 1$\,.  

The estimate Eq.(\ref{s4-pismall}) follows from considering the
bosonic Green's function  
\begin{equation}  \label{s4-pidef}
  \widetilde{\Pi}(\tau-\tau') = 
    \langle {\cal T}\{
      (f_\uparrow f_\downarrow)(\tau) \,
      (f^\dagger_\downarrow f^\dagger_\uparrow)(\tau') 
      \} \widetilde{\rangle}
    \;\,. 
\end{equation}
It acquires the form Eq.(\ref{s4-pifull}) when its diagrammatic
 is considered in SCTA for the Kondo
regime. In a fashion similar to the argument following
Eq.(\ref{s4-mixsmall}) we relate this propagator to a suitably chosen
thermal expectation value, namely 
\begin{displaymath}
  \langle{ (1-n^f_\uparrow)(1-n^f_\downarrow) \rangle} = 
    (\widetilde{ Z}/Z)\,
    \widetilde{\Pi}(\,(\tau-\tau')=0_+)
    \;\,.
\end{displaymath}
With Eq.(\ref{s4-kappa}) this results in 
\begin{displaymath}
  \int d \varepsilon\,
    \widetilde{\rho}^\Pi(\varepsilon)
    [-1-g(\varepsilon)]
    = \widetilde{\Pi}(0_+)
    \sim (\kappa)^1
    \;,\; 
\end{displaymath}
and Eq.(\ref{s4-pismall}) follows since the integrand is
always positive\cite{pap0995-note8}\,. In Appendix \ref{app-defect} this 
conclusion is drawn directly from Eq.(\ref{s4-pifull})\,. Similar to
$\widetilde{ \rho}^f$ the spectrum
$\widetilde{\rho}^\Pi(\omega)$ shows a threshold
behavior at $\omega=E^{Th}\approx \varepsilon^f+T_K$\,, as is shown
in Appendix \ref{app-sumrules}\,. At finite temperature this also implies
Eq.(\ref{s4-pismall})\,, following the arguments below
Eq.(\ref{s4-rhosmall})\,. 

The self energy $\Sigma^{ff}$ is given through the first term of
Eq.(\ref{s4-sigff2}) which reads after continuation to complex half planes
$i\omega_l\to z$\,, $\mbox{Im}(z)\ne 0$\,, 
\begin{mathletters}  \label{s4-tma-all}
  \begin{equation}  \label{s4-tma-sigf}
    \Sigma^{ff}(z) = 
      V^2\int d \varepsilon \,
      \rho^c(\varepsilon)[1-f(\varepsilon)]\,
      \widetilde{\Pi}(z + \varepsilon) \;\,.
  \end{equation}
This has to be inserted into $\widetilde{ G }^f$ given through
Eqs.(\ref{s4-skff}) and (\ref{s2-gtil0})\,,
  \begin{equation}  \label{s4-tma-gtil}
    \widetilde{ G}^f(z) = 
      \left[ z-\varepsilon^f-\Sigma^{ff}(z) \right]^{-1}
      \;\,. 
  \end{equation}
The right hand side of Eq.(\ref{s4-tma-sigf}) requires
  \begin{equation}  \label{s4-tma-pifull}
    \widetilde{\Pi}(z) = 
      \left[ \widetilde{\pi}(z)^{-1} 
                - \widetilde{ \sigma}(z) \right]^{-1}
      \;,\; 
  \end{equation}
according to the definition Eq.(\ref{s4-pifull})\,. $\widetilde{\pi}$ and
$\widetilde{ \sigma}$ given in Fig.\ \ref{fig-ladders} complete the
set of equations,
  \begin{equation}  \label{s4-tma-sigpi}
    \widetilde{ \sigma}(z) = 
      N_J V^2\int d \varepsilon\,
      \rho^c(\varepsilon)f(\varepsilon)\,
      \widetilde{ G}^f(z-\varepsilon)
      \;,\; 
  \end{equation}
with spin-degeneracy $N_J\equiv 2$\,, and
  \begin{equation}  \label{s4-tma-pi}
    \widetilde{\pi}(z) = 
      \int d \varepsilon\,
      \widetilde{\rho}^f(\varepsilon)
      [2f(\varepsilon) - 1] \,
      \widetilde{ G}^f(z-\varepsilon)
      \;\,. 
  \end{equation}
\end{mathletters}
The latter two expressions follow in leading order $\sim(\kappa)^0$ from
Eqs.(\ref{app-def-org}) derived in Appendix \ref{app-defect}\,. 
The self-consistency equations (\ref{s4-tma-all}) represent the 
Self-Consistent T-Approximation 
to the spin-degenerate $U\to\infty$ Anderson model in the Kondo
regime. Parameters are the 
$f$-level $\varepsilon^f$\,, hybridization matrix-element $V$\,, the
conduction band's density of states $\rho^c(\varepsilon)$\,, and the
temperature via the Fermi function $f(\varepsilon)$\,. 
The Green's function $\widetilde{ G}^f$\,, or its
spectrum $\widetilde{\rho}^f(\varepsilon)$ is by construction the
independent variable to be determined through a (numerical)
solution of Eqs.(\ref{s4-tma-all})\,. 
%
%%%*******************
%
\subsection{Physical \protect$f\protect$-Spectrum, Sum Rules}
%
%%%*******************
%
%
For calculating the $f$-excitation spectrum we now turn to
$\Sigma^{cc}$\,: From inspection of Fig.\ \ref{fig-kondself} it follows
\begin{eqnarray*}
  \lefteqn{ \Sigma^{cc}(i\omega_l) = }
    \\ 
  & & 
    V^2\int\!\!\!\int d \varepsilon d \varepsilon'\,
    \widetilde{\rho}^f(\varepsilon)
    \widetilde{\rho}^\Pi(\varepsilon')
    \frac{[1-f(\varepsilon)] + [-1-g(\varepsilon')]}
         {i\omega_l + \varepsilon - \varepsilon' }
    \;,\; 
\end{eqnarray*}
quite analogously to Eq.(\ref{s4-sigff2})\,. Here, with Eqs.(\ref{s4-rhosmall})
and (\ref{s4-pismall}) both terms of the right hand side are 
$\sim(\kappa)^1$\,, and the result fulfills
the general property Eq.(\ref{s4-orders})\,. With
Eq.(\ref{s4-fgf4}) the $f$-Green's function follows immediately, and
the $f$-excitation spectrum reads
\begin{eqnarray}
  \rho^f(\omega) & = &  \label{s4-rhof}
    -{\textstyle\frac{1}{\pi}}\mbox{Im}\,F(\omega+i0_+)
    \\ 
  & = &  \nonumber 
    \int\! d \varepsilon
    \left[ 
    \overline{\overline{\rho}}^f(\varepsilon)
    \widetilde{\rho}^\Pi(\omega + \varepsilon ) + 
%%%         \right.
%%%    \\ \nonumber 
%%%  & & \hspace*{ 4mm}
%%%    \left. + 
    \widetilde{\rho}^f(\varepsilon)
    \overline{\overline{\rho}}^\Pi(\omega + \varepsilon )
         \right]
\end{eqnarray}
Projected Spectra have been introduced here, which are
$\sim(\kappa)^0$ and positive for all energies, 
\begin{mathletters}  \label{s4-aux-all}
  \begin{eqnarray}
    \label{s4-aux-f}
    \overline{\overline{\rho}}^f(\varepsilon) & = &  
      \frac{\widetilde{ Z}}{Z}\,
      \widetilde{\rho}^f(\varepsilon)[1-f(\varepsilon)]
      \;,\;
      \\ 
    \label{s4-aux-pi}
    \overline{\overline{\rho}}^\Pi(\varepsilon) & = &  
      \frac{\widetilde{ Z}}{Z}\,
      \widetilde{\rho}^\Pi(\varepsilon)[-1-g(\varepsilon)]
      \;\,. 
  \end{eqnarray}
\end{mathletters}
It is suitable to use an extra set of equations for 
$\overline{\overline{\rho}}^f$ and $\overline{\overline{\rho}}^\Pi$\,,
in analogy to the Defect Equations
in NCA-theory \cite{note-bic}\,. These are derived in Appendix
\ref{app-defect} and read 
\begin{mathletters}  \label{s4-def-all}
  \begin{equation}
    \label{s4-def-f}
    \overline{\overline{\rho}}^f(\omega) = 
      \left| \widetilde{ G}^f(\omega) \right|^2 
      V^2 \int d \varepsilon\,
      \rho^c(\varepsilon) f(\varepsilon) 
      \,\overline{\overline{\rho}}^\Pi(\omega+\varepsilon)
      \;,\; 
  \end{equation}
  \begin{equation}
    \label{s4-def-pi}
    \overline{\overline{\rho}}^\Pi(\omega) = 
      \left| \widetilde{\Pi}(\omega) \right|^2 
      N_J V^2 \int d \varepsilon\,
      \rho^c(\varepsilon) [1-f(\varepsilon)] 
      \,\overline{\overline{\rho}}^f(\omega-\varepsilon)
      \;,\;
  \end{equation}
\end{mathletters}
with fixed spin degeneracy $N_J\equiv 2$\,. 

For the spectra occurring in Eqs.(\ref{s4-tma-all}) and (\ref{s4-rhof}),
(\ref{s4-def-all}) a set of exact sum rules is proven in Appendix
\ref{app-sumrules}\,. Here we quote the results: The integrated
spectral weight of $\widetilde{ \rho}^f$ and $\widetilde{ \rho}^\Pi$ is
given by
\begin{mathletters}   \label{s4-sum-all}
  \begin{eqnarray}
    \int d\varepsilon\,    \label{s4-sum-ftil}
      \widetilde{ \rho}^f(\varepsilon) & = & 1
      \;,\;
      \\ 
    \int d\varepsilon\,    \label{s4-sum-pi}
      \widetilde{ \rho}^\Pi(\varepsilon) & = & 
      1 - 2\int d\varepsilon\,
      \widetilde{ \rho}^f(\varepsilon)
      [1 - f(\varepsilon)]
      \;,\;
  \end{eqnarray}
with Eq.(\ref{s4-ffsmall}) the weight of $\widetilde{\rho}^\Pi$ can be
set equal to one. Projected Spectra are related via 
  \begin{equation}    \label{s4-sum-defect}
    \int d\varepsilon\,
      \left[ 2 \overline{\overline{\rho}}^f(\varepsilon) - 
             \overline{\overline{\rho}}^\Pi(\varepsilon) \right]
      = 1
      \;,\;
  \end{equation}
and the physical $f$-spectrum is normalized according to
  \begin{equation}    \label{s4-sum-fnorm}
    \int d\varepsilon\,
      \rho^f(\varepsilon) = 
      \langle 1-n^f_{\uparrow,\downarrow} \rangle = 
      \int d\varepsilon\,
      \overline{\overline{\rho}}^f(\varepsilon)
      \;\,.
  \end{equation}
\end{mathletters}
Note that the integrated $f$-electron spectral weight is
not equal to unity, since excitations at high energies $\sim U$ are
excluded here; compare to Section \ref{sec-wick}\,. 

In Appendix \ref{app-sumrules} the SCTA is tested against sum rules. It
turns out that Eqs.(\ref{s4-sum-ftil}) and (\ref{s4-sum-pi}) are
strictly fulfilled by the self-consistency equations
(\ref{s4-tma-all})\,, whereas the total $f$-spectral weight comes out
slightly too large,
$    \int d\varepsilon\,
    \rho^f(\varepsilon) = 
    3 \int d\varepsilon\,
      \overline{\overline{\rho}}^f(\varepsilon) - 1 $\,. 
Nevertheless, the numerical difference to Eq.(\ref{s4-sum-fnorm}) is
small in the deep Kondo regime, where
$\langle n^f_\sigma \rangle \lesssim 1/2$\,. 
%
%
%
%%%******************* section starts here ***
\section{Numerical Results and Relation to The NCA}
\label{sec-numeric}
This Section is devoted to numerical results and their interpretation.
In addition connections of the SCTA with the Non-Crossing Approximation
\cite{note-bic} (NCA) are discussed.

The SCTA-Equations (\ref{s4-tma-all}) have been solved numerically
at the real axis $z\to\omega+i0_+$ through iteration of
$\widetilde{\rho}^f(\omega)$\,. After convergence, the proper normalization
of $\widetilde{\rho}^f$ and $\widetilde{ \rho}^\Pi$ has been checked. 
Results shown here correspond to a flat conduction-band DOS 
$\rho^c(\omega) = \Theta(D-|\omega|) / 2D$
with cut-off $D$ and 
parameters $D=10.0\Delta$\,, $\varepsilon^f=-3.0\Delta$\,, $N_J\equiv
2$ in the Kondo regime, where $\Delta=\pi V^2\rho^c(0)$ denotes the energy
scale of the non-interacting impurity model. The Kondo energy
\cite{kriwilwil80b,mue84} takes a value $T_K\approx
0.016\Delta$\,. In Fig.\ \ref{fig-spectra} the spectra 
$\widetilde{\rho}^f$ and $\widetilde{\rho}^\Pi$ are 
displayed for a temperature
$k_BT=0.1T_K$\,. A threshold behavior, which has been shown above to
be a feature of these spectra at $T\to 0$ (see also Appendix
\ref{app-sumrules}) shows up in the form of a sharp resonance near
the bare $f$-level $\varepsilon^f$\,. The Projected Spectra
$\overline{\overline{\rho}}^f$\,, 
$\overline{\overline{\rho}}^\Pi$ are obtained by iteration of
Eqs.(\ref{s4-def-all}) and have been normalized according to
Eq.(\ref{s4-sum-defect}) (a common 
prefactor is arbitrary in Eqs.(\ref{s4-def-all})). From Eq.(\ref{s4-rhof})
then follows the $f$-excitation spectrum $\rho^f$\,. Since it 
does not strictly fulfill its sum rule Eq.(\ref{s4-sum-fnorm})\,, the
latter has been used to enforce the correct norm of $\rho^f$\,. 

In Fig.\ \ref{fig-fdos} the calculated $f$-spectrum is shown, together
with an NCA result for the same set of parameters and temperature. It
shows the well known features, a `charge resonance' (CR) near the $f$-level
and a sharp Abrikosov--Suhl resonance (ASR) near the Fermi edge
$\omega=0$\,. In SCTA the CR is located
slightly above $\varepsilon^f=-3.0\Delta$\,,
which is in contradiction to intuition (the gain in kinetic energy due
to hybridization should {\em lower} the $f$-level) and the NCA
result. Obviously, in SCTA some spectral weight is transfered to
energies near the Fermi level, and the amplitude of the ASR is in
general too high. As a consequence local Fermi-liquid properties
\cite{yosyam75} and the Friedel--Langreth sum-rule
\cite{yoszaw82} are not fulfilled. The $f$-valence
$N_J\langle n^f_\sigma \rangle$ as derived from
Eq.(\ref{s4-sum-fnorm}) takes a value $\approx 0.66$ too small for the Kondo
regime. As an origin of the misplaced spectral weight in the
$f$-excitation spectrum $\rho^f$ we might take 
the representation of the $f$-Green's function in SCTA, i.e.\ the self 
energy $\Sigma^{cc}$ as displayed in Fig.\ \ref{fig-kondself} (bottom), since
it violates the normalization rule Eq.(\ref{s4-sum-fnorm})\,. On the other
hand, the characteristic shape of the `charge resonance' is
already obvious in $\widetilde{ \rho}^\Pi$ shown in Fig.\
\ref{fig-spectra}\,, which results solely from $\Sigma^{ff}$ (depicted in
Fig.\ \ref{fig-kondself} top) and enters $\rho^f$ through the convolution in
Eq.(\ref{s4-rhof})\,. Also the $f$-valence is computed    
directly from $\overline{\overline{\rho}}^f$ via the 
sum rule Eq.(\ref{s4-sum-fnorm}) and does not involve $\Sigma^{cc}$ either. 

Some insight may be gained from a certain non-systematic limit of the SCTA,
which reproduces the NCA\,: 
If the ladder structure in $\Sigma^{ff}$ or $\Sigma^{cc}$ (see Fig.\
\ref{fig-kondself}) is read from the left to the right, the `ladder
element' $\widetilde{ \pi}$ (displayed in Fig.\ \ref{fig-ladders} left)
could be viewed as two holes of opposite spin coexisting in 
the $f$-orbital. This empty-orbital state then would be renormalized through 
intermediate particle--hole excited states with single $f$-occupancy,   
represented by $\widetilde{\sigma}$ (see Fig.\ \ref{fig-ladders} right). 
Following this interpretation, our bosonic Green's function
$\widetilde{\Pi}(z)$ appears similar to the NCA-propagator 
$G_0(z)$ for the empty $f$-orbital \cite{biccoxwil87}\,. In fact, if the
renormalization of lines within $\widetilde{ \pi}$ is ignored (NCA Limit),
\begin{displaymath}
  \widetilde{ \pi}(z) \to
    \widetilde{ \pi}^{NCA}(z) :=
    1 / [ z - 2 \varepsilon^f ]
    \;,\;
\end{displaymath}
the NCA equations \cite{note-bic} are reproduced after a re-definition of 
$z \to z'=2 \varepsilon^f - z$\,.
Although the interpretation outlined above seems obvious, it may only
serve as an instructive starting point, because the Green's function
$\widetilde{ G}^f$ (the 
dashed double line in diagrams) does {\em not} represent any physical
particle. The `formal meaning' of both $\widetilde{ \Pi}$ and
$\widetilde{ G}^f$ has to be viewed quite analogously to
unphysical propagators $G_0$ and 
$G_1$ of Resolvent-Perturbation Theory. In the limit
of zero temperature this analogy even turns to an equality, as is
argued in Appendix \ref{app-sumrules}\,.
Nevertheless, in general both approaches are quite
different, which is apparent if the definitions of
Green's functions and spectra in Eqs.(\ref{s3-gtilff}),
(\ref{s4-pidef}), (\ref{s4-aux-all}) and sum rules
Eqs.(\ref{s4-sum-all}) are compared to their respective counterparts of  the
Resolvent Method (see e.g.\ Ref.\ \onlinecite{biccoxwil87}). Most
striking are the completely different diagram rules and the explicit
use of the Kondo regime within the approach presented here.

Utilizing the NCA Limit  the unexpected shape of
the `charge resonance' (CR) in SCTA is traced back to the 
self energy $\Sigma^{ff}$\,: In the NCA Limit
the `ladder element' $\widetilde{\pi}$ shows an imaginary part
$-\mbox{Im}\,\widetilde{\pi}^{NCA}(\omega+i0_+) = 
  \pi\delta(\omega-2 \varepsilon^f)$
sharply peaked around $2 \varepsilon^f$\,, which leads to the NCA-curve
in Fig.\ \ref{fig-fdos}\,. 
In going to the ladder approximation, this is replaced by
$-\mbox{Im}\,\widetilde{\pi}(\omega+i0_+) = 
  \pi \int d \varepsilon\,
  \widetilde{\rho}^f(\varepsilon)
  \widetilde{\rho}^f(\omega-\varepsilon)$\,.
Accordingly $-\mbox{Im}\,\widetilde{\pi}$ has a 
broadened shape like $\widetilde{\rho}^f$  displayed in Fig.\
\ref{fig-spectra} (full line), with the peak located
above $2\varepsilon^f$\,. Hence the
CR in $\rho^f$ is shifted to higher frequencies and gets the asymmetric 
shape visible in the SCTA-curve in Fig.\ \ref{fig-fdos}\,. 
A few more conclusions are drawn in the Summary. 
%
%
%
%
%%%******************* section starts here ***
\section{Application to The Lattice Model}
\label{sec-lattice}
\subsection{Elementary Perturbation Expansion}
\label{sec-lattice-direct}
The strong-coupling-perturbation
expansion for the Anderson-lattice model, as the generic model for e.g.\
Heavy-Fermion Systems is discussed from the Feynman-diagram
technique's point of view. The main difficulty encountered when
performing an expansion around the Atomic Limit of lattice models is
the so-called Excluded-Volume Problem \cite{bro61} (EVP), which
hinders the application of the Linked-Cluster Theorem: Within the
Resolvent-Perturbation formalism all parts of a lattice-diagram which
involve a certain $f$-orbital site $R_i\equiv i$ are necessarily
linked together \cite{grekei81}\,. Accordingly 
diagrams have to be broken up (`decoupled') into unlinked diagrams
\cite{gre84,kur85} in order to obtain an expansion of thermal densities 
like the (site dependent) lattice $f$-Green's function $F^{lat}_{ij}$
through connected diagrams, leading to the known LNCA
\cite{gre87,greprukei88} and XNCA 
\cite{kimkurkas90} schemes. Specifying quasi particle
interactions (cumulants) consistently remains a difficult task
\cite{gre87}\,. We consider here the 
lattice model in a fashion analogous to the impurity treatment in Sect.\
\ref{sec-wickwick}\,. It shall turn out, although Feynman-diagrams are
obtained, that the EVP cannot be avoided completely; nevertheless it
takes a much `weaker' form than in Resolvent-Perturbation Theory.

After having decoupled states with at least one doubly occupied
$f$-orbital from those with no double $f$-occupancy via a canonical 
transformation of the Anderson-lattice Hamiltonian in leading order
$V/U$\,, we arrive at an expression for $F^{lat}_{ij,\sigma}$ similar to
Eq.(\ref{s2-fgf2})\,. For large $U$\,, i.e.\ $J=0$ it reads in the
interaction picture
\begin{eqnarray}   \label{s6-fgf}
  \lefteqn{ F^{lat}_{ij,\sigma}(\tau,\tau') = 
    - \left( \widetilde{ Z}^{lat(0)}/Z^{lat} \right)  }
    \\ 
  & &  \nonumber
    \times\,
    \Bigl\langle {\cal T}\Bigl\{
    \exp\left(-\int_0^\beta d\overline{\tau}
              \sum_i H^V_{01;i}(\overline{\tau}) \right)
    \\ 
  & &  \nonumber
    \times\,
    (f_{i,-\sigma}f^\dagger_{i,-\sigma}f_{i \sigma})(\tau)
    (f_{j,-\sigma}f^\dagger_{j,-\sigma}f^\dagger_{j \sigma})(\tau')
    \Bigr\}
    \\ 
  & &  \nonumber
    \times\,
    \prod_{\mu\ne i,j}
    e^{-\beta U n^f_{\mu\uparrow} n^f_{\mu\downarrow}}
    \Bigr\rangle^{\widetilde{lat}(0)}
\end{eqnarray}
with 
$\widetilde{ Z}^{lat(0)}= 
   \mbox{Tr}[\exp(-\beta \widetilde{ H}^{lat(0)})]$\,.
The unperturbed (atomic) lattice Hamiltonian is split into the bilinear part
\begin{displaymath}
  \widetilde{ H}^{lat(0)} = 
    \sum_{k,\sigma}\varepsilon_k c^\dagger_{k \sigma}c_{k \sigma} +
    \sum_{i,\sigma}\varepsilon^f_\sigma f^\dagger_{i \sigma}f_{i \sigma}
\end{displaymath}
and the local $f$-repulsion $\sim U$\,. The latter does not influence
the time dependent operators occurring in Eq.(\ref{s6-fgf}) and appears
only in the product of
$e$-factors. The perturbations $H^V_{01;i}$ are
generalizations of Eq.(\ref{s2-vpart01}) for each site
$i=1,\ldots,N_{BZ}$\,. Due to the fact that states with double local
$f$-occupancy do contribute to the trace for all lattice sites $\mu$
different from $i$ and $j$\,, the $e$-factors in general cannot be set
equal to unity, in contrast to the 
single site $i=j\equiv 1$ problem. The contribution of order $V^n$
to the perturbation series arising from Eq.(\ref{s6-fgf}) is proportional to
\begin{equation}  \label{s6-contrib}
 \begin{array}{c}  \displaystyle 
  \langle {\cal T}\{
    H^V_{01;i_1} \ldots H^V_{01;i_n} 
    \cdot(..i..)(\tau)\cdot(..j..)(\tau') \}
    \rangle^{\widetilde{ lat}(0)}
    \times 
    \\  \displaystyle 
  \times\,
    \prod_{\mu\not\in L}
    \langle e^{-\beta U n^f_{\mu\uparrow} n^f_{\mu\downarrow}}
    \rangle^{\widetilde{ lat}(0)}
    \;,\; 
 \end{array}
\end{equation}
where time variables to be integrated out are omitted. Site indices
$i_1,\ldots,i_n; i,j$ are kept fixed for the moment and are assumed to
cover a certain set $L=\{ \nu_1,\nu_2,\ldots,\nu_l \}$ of $l\le (n+2)$
{\em mutual distinct} lattice sites. Accordingly the $e$-factors from
Eq.(\ref{s6-fgf}) for all
sites $\overline{\mu}\in L$ act as unity since double $f$-occupancy is
projected out by $H^V_{01;\overline{\mu}}$\,, and the expression takes
the factorized form Eq.(\ref{s6-contrib})\,. Wick's theorem is
now applicable to the first term in Eq.(\ref{s6-contrib})\,, whereas
the second term yields an additional $c$-number,
$\prod_{\mu\not\in L}
   \langle \ldots \rangle^{\widetilde{ lat}(0)} = 
   (\sum_\sigma e^{\beta \varepsilon^f_\sigma})^{(n-l)}$
for $k_BT\ll|\varepsilon^f|$\,. That is, a bookkeeping of the {\em
number $l$ of distinct sites} involved has to take place when sums over
site indices are performed. This `weak form' of the Excluded-Volume
Problem prevents the conventional direct cancelation of disconnected
Feynman diagrams 
in the lattice $f$-Green's function. Nevertheless an expansion in
Feynman diagrams is gained, i.e.\ any kind of `decoupling' of lattice
diagrams becomes unnecessary. Possible applications will be indicated
in Sect.\ \ref{sec-summa}\,. 
\subsection{Local Approximation}
\label{sec-lattice-local}
A from the outset more approximative approach to lattice
models is the well known Local Approximation (also referred to as
Dynamical Mean-Field Theory), which emerges from 
considerations of the limit of infinite spatial dimension
\cite{metvol89,mue89} ($d\to\infty$) via weak-coupling-perturbational
and -functional methods (see e.g.\ Ref.\
\onlinecite{kurwat87,schczy89,bramil90,donvol90,jan91,jar92,geokotsi92})
as well as 
strong-coupling-perturbation theories \cite{met91,barcha93,jbepl} doing
with Hubbard cumulants \cite{hub66}\,. For the Anderson-lattice
model the different $d\to\infty$-approaches \cite{schczy89,hueqin94,jbepl}
lead to identical results \cite{jbepl,kei95}\,, which 
becomes obvious if
self-energies are eliminated from equations in Refs.\
\onlinecite{schczy89,hueqin94}\,. The lattice
$f$-Green's function $F^{lat}_{k \sigma}$ in $k$-space turns out
equivalent to the XNCA expression \cite{kimkurkas90}\,, 
\begin{displaymath}
  F^{lat}_{k \sigma}(z) = 
    \left[ (F_\sigma(z))^{-1} - V^2 \left(
           \frac{1}{z-\varepsilon_k} - {\cal G}^c_\sigma(z) \right)
    \right]^{-1}
\end{displaymath}
with an effective impurity (EI) model's conduction-electron Green's
function ${\cal G}^c_\sigma$ determined through 
the lattice-self-consistency condition (Lattice-SC)\,,
$\frac{1}{N_{BZ}}\sum_k F^{lat}_{k \sigma}(z) = 
   F_\sigma(z)$\,.
The EI model $f$-Green's function is designated as $F_\sigma$\,. In calculating
$F_\sigma$  numerically via the SCTA theory within a loop that iterates
${\cal G}^c_\sigma$ to convergence, the approach described in previous
Sections is extended to the $U\to\infty$ 
Anderson-lattice model. Unfortunately we encounter a difficulty
well known from the XNCA \cite{kimkurkas90}\,: Except at
extremely high temperatures the spectrum of ${\cal G}^c_\sigma$
develops a peak structure near the Fermi level, indicating an
enhancement of the Kondo effect \cite{kimkurkas87}\,; but if ${\cal
G}^c_\sigma$ is re-inserted into the SCTA\,, the spectrum of $F_\sigma$
does not alter significantly, i.e.\ it does not show any tendency toward
a hybridization gap in the Abrikosov--Suhl resonance. Consequently the
Lattice-SC cannot be fulfilled. Comments on this result are given in
the following Summary.  
%
%
%
%%%******************* section starts here ***
\section{Summary and Conclusions}
\label{sec-summa}
A detailed description of a self-consistent
strong-coupling-perturbation theory for the Anderson model has been
given, which extends a Feynman-diagram technique proposed recently
\cite{jbepl,jbsces}\,. It was shown in Sect.\ \ref{sec-wickwick} that
Wick's theorem can be applied 
{\em directly} within a perturbation expansion in the hybridization
$V$\,, i.e.\ the coupling of a localized orbital ($f$-orbital) to
delocalized conduction-electron states. Although the strong Coulomb
interaction $U$ of electrons on the 
$f$-orbital is included in the unperturbed Atomic Limit, conventional
Feynman-diagram rules are obtained in the sequel, involving the
two-particle vertices $\sim V$ and $\sim V^2/U$ shown in Fig.\
\ref{fig-vertgf}\,. This kind of Feynman-perturbation theory starts
from a 
canonical transformation of the Hamiltonian, which eliminates in
leading order $V/U$ the charge fluctuations into states with double
$f$-occupancy. It has also been pointed out in Sect.\
\ref{sec-wickwick} that an application to the (multi channel) Kondo
model is apparent. 

Based on the diagram rules for 
$U\to\infty$ the Self-Consistent T-Approximation (SCTA) has been
derived in Sect.\ \ref{sec-general}\,. By means of
irreducible vertex functions compliant to symmetry relations
(Crossing Relations) a matrix-type self energy
\mbox{\boldmath$\Sigma$} has been set up; it is displayed in Figs.\
\ref{fig-selfbethe}\,, 
\ref{fig-selftma}\,. This skeleton-self energy constitutes a $\Phi$-derivable
approximation for a set of unphysical propagators as well as 
a  prescription for constructing the physical $f$-Green's
function from 
these propagators. It contains ladder sums of the diagram elements
shown in Fig.\ \ref{fig-logloops} to infinite order $V$\,; the
latter have been identified as important for the formation of the Kondo
effect. A version of the SCTA 
suitable for the Kondo regime emerged in Sect.\ \ref{sec-kondo} from
an expansion of \mbox{\boldmath$\Sigma$} in a
non-perturbative small parameter. Only one of the unphysical Green's
functions remains renormalized in this case,
represented through the self-consistent SCTA-Equations
(\ref{s4-tma-all})\,. Some relations to the Non-Crossing Approximation
(NCA) were explored in Sect.\ \ref{sec-numeric}\,. Furthermore the
SCTA contains the NCA as a certain non-systematic limit. 

The numerical results for the physical $f$-Green's function 
presented in Sect.\ \ref{sec-numeric} reproduce the known 
correlation-induced features of the $f$-excitation spectrum. 
Concerning Fermi-liquid properties, it became apparent that the SCTA
does not furnish results which improve those of the NCA\,. 
Nevertheless, the formulation of enhanced approximations will be
straight forward since standard techniques for Feynman diagrams (e.g.\
Parquet equations) can be used, and the expansion
for the Kondo regime is not restricted to specific diagram classes 
(see Appendix \ref{app-loop}). Going beyond the summation
of ladder diagrams in the $\Phi$-functional will involve `true'
retarded vertex functions. The importance of vertex corrections to the
NCA has already been pointed out \cite{mue84,and95,coskrowoe96}\,. Here 
vertex corrections to the matrix-type self energy
\mbox{\boldmath$\Sigma$} yield a consistent modification of the 
self-consistency equations as well as the representation of the physical
$f$-Green's function. Actually both have significant
influence on the $f$-spectrum and require corrections, as has been
argued in Sect.\ 
\ref{sec-numeric}\,. Further studies based on the technique developed here
might therefore be fruitful with regard to dynamical properties of the
spin-degenerate model in the problematic region of very low temperature. 

Two possible ways of extending the technique to the ($U\to\infty$)
Anderson-lattice model, an elementary perturbation expansion on 
the lattice and the 
application within the Local Approximation, have been discussed in Sects.\
\ref{sec-lattice-direct} and \ref{sec-lattice-local} respectively. 
For the elementary expansion it turned out that any contribution to
the $f$-Green's function is indeed decomposed into Feynman diagrams by
virtue of Wick's theorem,  
but in addition the number of mutual different $f$-orbital
sites actually involved enters through a $c$-number 
(`weak' Excluded-Volume Problem). Thus an
expansion in connected diagrams may not obtained in a simple fashion.
Nevertheless, the definition of quasi particles and their interactions
seems manageable. Accordingly the approach looks
quite suitable for models with a small number of impurities with local
interaction, e.g.\ the two-impurity multi-channel Kondo model
\cite{ingjonwil92}\,.  Concerning the Anderson-lattice and 
$t$--$J$-model we believe that our method will also show some impact,
especially 
on the study of non-local spin fluctuations via strong-coupling theory
for e.g.\ High-T$_c$-Superconductors \cite{bat94,shedes95rev}
or Heavy-Fermion Systems near a quantum-phase transition 
\cite{steetal94,mapetal94}\,. 

In the Local Approximation (or Dynamical Mean-Field Theory, reviewed
briefly in Sect.\ 
\ref{sec-lattice-local}) the $f$-Green's function of the lattice model
is given through that of an effective impurity (EI) 
model subject to a lattice-self-consistency condition (Lattice-SC). 
The SCTA equations from Sect.\ \ref{sec-kondo} were used to solve the
EI model, but convergence to the Lattice-SC has not been achieved; a
feature already known from the NCA within Local Approximation
(i.e.\ the XNCA) \cite{kei95}\,. Such a behavior
is not found if the Lattice-SC for the Anderson-lattice model is 
combined with methods different from strong-coupling perturbation theory 
(e.g.\ QMC \cite{jarakhpru93,jar95}\,, numerical
diagonalization \cite{roz95} or weak-coupling perturbation theory
\cite{muthir94,schczy90}), or if strong-coupling theory is used within a
different Lattice-SC for the same model, as is the case of the LNCA 
\cite{greprukei88,kei95}\,. In order to clarify this inconsistent
situation we propose considering alternative 
strong-coupling expansions for the lattice model, which do not rely on
the Local Approximation or the Resolvent Method. The elementary expansion 
outlined in Sect.\ \ref{sec-lattice-direct} may serve as a starting
point. 
%
%
%
%%%************************
\section*{Acknowledgments}
The author wishes to thank Prof.~\mbox{N.~Grewe} for many valuable
discussions on the subject of this work and support during preparation
of the manuscript. A critical reading of the manuscript by Drs.\ 
\mbox{F.~Anders} and \mbox{Th.~Pruschke} has been greatly
appreciated. Useful conversation to Profs.\ \mbox{P.~W{\"o}lfle} and
\mbox{J.~Keller}, and Drs.\ \mbox{T.~A.~Costi} and \mbox{T.~Kopp} is
also gratefully acknowledged. 
\appendix
%
%%%**************** Appendix starts here ***
\section{Effective Hamiltonian}
\label{app-cano}

An effective Hamiltonian $H''$\,, where the mixing $H^V_{12}$
involving local double occupation is removed to first order $V/U$ is
derived from Eq.(\ref{s2-hamx}) in two steps:
The first unitary transformation \cite{schwol66,chaspaole77} is 
$H \to H'$\,,
\begin{displaymath}
  H' = 
    e^{iS} H e^{-iS} = 
    H + [iS,H] + \case{1}{2}[iS,[iS,H]] + \ldots
\end{displaymath}
with a generator $iS$ subject to the requirement 
\begin{displaymath}
  [iS,(H^c + H^{loc})] = - H^V_{12} \;\,.
\end{displaymath}
An appropriate choice is 
\begin{equation}  \label{app-cano-gen1}
  iS = 
    \frac{1}{\sqrt{N_{BZ}}} \sum_{k,\sigma}
    {\cal J}_\sigma(k)\,
    n^f_{-\sigma}f^\dagger_\sigma c_{k \sigma} - h.c.
\end{equation}
with 
${\cal J}_\sigma(k) = V_k / (\varepsilon^f_\sigma+U-\varepsilon_{k \sigma})$
\,. Since $H^V_{01}$ is kept in $H'$\,, additional operators mixing
Hilbert space sectors ${\cal H}_{0,1}$ and ${\cal H}_{2}$ emerge
through 
$[iS,H^V_{01}]$\,, which is of order $V/U$\,. To this order they
vanish in a second transformation 
\begin{displaymath}
  H' \to H'' = 
    H' + [iS',H'] + \ldots
\end{displaymath}
using a generator $iS'$ which fulfills 
\begin{displaymath}
  [iS',(H^c + H^{loc})] = - [iS,H^V_{01}]
    \;\,. 
\end{displaymath}
It is explicitly given by
\begin{displaymath}
  iS' = 
    \frac{1}{N_{BZ}} \sum_{k,q,\sigma}
    {\cal K}_\sigma(k,q)\,
    f^\dagger_\sigma f^\dagger_{-\sigma}c_{q,-\sigma}c_{k \sigma}
    - h.c.
\end{displaymath}
involving coefficients ${\cal K}_\sigma(k,q)$ of order $V^2/U^2$\,. 
The resulting effective Hamiltonian $H''$ reads to first order $V/U$\,,
\begin{displaymath}
  H'' = 
    H^c + H^{loc} + H^V_{01} + H^J
    \;,\; 
\end{displaymath}
where the Coulomb repulsion $U$ in $H^{loc}$ is slightly shifted,
\begin{displaymath}
  U \to U' = U + \frac{1}{N_{BZ}} \sum_{k,\sigma}
    \frac{|V_k|^2}{\varepsilon^f_\sigma + U - \varepsilon_{k \sigma}}
\end{displaymath}
and a spin--spin interaction emerges, which is part of 
\begin{eqnarray*}
  H^J & = & 
    \frac{1}{N_{BZ}} \sum_{k,q,\sigma}
    \frac{1}{2}(-{\cal J}_\sigma(k)\,V^\ast_q)\cdot
    \\ \\
  & & \;\;\cdot
    \left( f^\dagger_\sigma c^\dagger_{q,-\sigma}
           f_{-\sigma} c_{k \sigma} + 
           f^\dagger_\sigma c^\dagger_{q,-\sigma}
           c_{k,-\sigma} f_\sigma \right) + h.c.
\end{eqnarray*}
The exchange integral originating from Eq.(\ref{app-cano-gen1})
is simplified for large $U$ and $V_k=V$ to 
\begin{displaymath}
  {\cal J}_\sigma(k)V^\ast_q \approx |V|^2/(\varepsilon^f+U)\equiv -J_U
  \;,\; 
\end{displaymath}
and $H^J$ takes the form Eq.(\ref{s2-exham})\,. 
%
%
%
%%%**************** Appendix starts here ***
\section{The Loop Theorem}
\label{app-loop}
In Section \ref{sec-kondo} counting of a certain kind of loops is used
to separate out important self-energy diagrams for the Kondo regime. The
underlying `loop theorem' for skeleton diagrams is derived as follows. 
Consider a {\em closed path} in an arbitrary diagram, as introduced in Section
\ref{sec-kondo} below Eq.(\ref{s4-rhosmall})\,: It consists solely of 
equally directed renormalized Green`s functions $\widetilde{ G}^f$
(dashed double lines, see Fig.\ \ref{fig-matrix}) and vertices as displayed
in Fig.\ \ref{fig-vertgf}\,(a)\,. Vertices may be passed from any
incoming dashed line to any outgoing dashed line, i.e.\ the path may
involve both spin directions. None of the diagram's 
lines and vertices is to be
touched more than once. In Fig.\ \ref{fig-omitted}~(a) some examples
are shown. 
The analytical contribution from a {\em closed path} running on $k$ dashed
double lines involves also $k$ distinct vertices and is proportional to
\begin{eqnarray}
  \lefteqn{ L^{(k)} = }   \label{app-loop-loop}
    \\ 
  \;\;\; & &  \nonumber 
    \pm \frac{1}{\beta} \sum_{\omega_n} e^{-i\omega_n 0_+}
    \left[ \widetilde{ G}^f(i\omega_n - i\nu_1 ) \ldots 
           \widetilde{ G}^f(i\omega_n - i\nu_k )
    \right]
    \;\,. 
\end{eqnarray}
It is coupled to the remainder of the diagram through $k$ fixed bosonic
frequencies labeled $\nu_1, \ldots, \nu_k$\,. These are kept unequal in
pairs through suitably fixed infinitesimal increments \cite{keimor} 
$\nu_l\to\nu_l+\delta_l$\,. The sum over the free fermionic
loop-frequency $\omega_n$ is rewritten as usual \cite{negorl}\,, 
\begin{eqnarray*}
  L^{(k)} & = & 
    \mp \int d\varepsilon_1 \ldots \varepsilon_k \,
    \widetilde{ \rho}^f(\varepsilon_1) \ldots
    \widetilde{ \rho}^f(\varepsilon_k) 
    \\ 
  & & \hspace*{ 2mm} \times
    \oint\limits_\Gamma\frac{d z}{2\pi i}
    \frac{ [1-f(z)] }
         { (z-i\nu_1-\varepsilon_1) \ldots (z-i\nu_k-\varepsilon_k) }
    \;,\;
\end{eqnarray*}
where $\widetilde{ G}^f$ is expressed via its spectrum  
Eq.(\ref{s4-rhotildef})\,. The contour
$\Gamma$ encircles all zeros of the denominator. The term $[1-f(z)]$ is
a consequence of the exponential factor in Eq.(\ref{app-loop-loop})\,,
which keeps track of non-normal ordering of operators (ref.\ Sect.\
\ref{sec-wick}). Performing the contour integral leads to
\begin{displaymath}
  L^{(k)} = 
    \mp \sum_{l=1}^k \int d \varepsilon_l\,
    \widetilde{ \rho}^f(\varepsilon_l) [1-f(\varepsilon_l)] \,
    \prod_{l'\ne l} 
    \widetilde{ G}^f(\varepsilon_l+i\nu_l-i\nu_{l'})
\end{displaymath}
Now all other free frequencies in the diagram are integrated out one after the
other, and the diagram appears as a sum over $k$ terms.
These are regular for finite temperature, and
the infinitesimal increments introduced above become unnecessary. 
The function 
$\widetilde{ \rho}^f(\varepsilon)[1-f(\varepsilon)]\sim\kappa$
discussed in Section \ref{sec-kondo}
shows up in each term here and gives the diagram the small overall order of
magnitude 
$\kappa = \exp(\beta(\varepsilon^f+T_K))$
in the Kondo regime. The argument becomes apparently invalid if at least one
of the Green's functions $\widetilde{ G}^f$ on the path is reversed in
orientation, i.e.\ $\widetilde{ \rho}^f(\varepsilon) \to \widetilde{
\rho}^f(-\varepsilon)$\,, or is replaced by a conduction-electron
propagator $\widetilde{ \rho}^f(\varepsilon) \to \rho^c(\varepsilon)$\,. 
In a straight-forward generalization, the resulting `loop theorem' is
stated: $p$ non-overlapping {\em closed directed paths} in a skeleton
diagram are {\em sufficient} for an order of magnitude
$\sim(\kappa)^p$\,. 

As an application we determine the order of magnitude of self energies
$\Sigma^{a'a}, a'a=ff,cf,fc,cc$\,, without reference to any specific
approximation. The result has been quoted in
Eq.(\ref{s4-orders})\,. Consider at first a diagram to $\Sigma^{cc}$\,,
or more generally 
an arbitrary connected dressed skeleton diagram with one incoming and one
outgoing conduction-electron line $G^c$\,. It shows necessarily at least
one {\em closed path} as defined above: 
We construct a path running
solely on dashed lines $\widetilde{ G}^f$\,, starting on an outgoing
dashed line of an arbitrary chosen vertex within the diagram. We always
follow the direction of arrows, and no line or vertex is visited
twice. Owing to the fact that a vertex (displayed in Fig.\
\ref{fig-vertgf}\,(a) left or right) entered on the way can be left via at
least one dashed line, the path cannot terminate within the diagram. Therefore
(i) the procedure can lead back to the vertex we started from, and a
{\em closed path} is completed; (ii) a vertex with two incoming dashed
lines (Fig.\ \ref{fig-vertgf}\,(a) right) is re-entered for the first
time on its hitherto unused line, which constitutes a {\em closed
path}\/, too; (iii) the path we are constructing leaves the diagram via an
external outgoing dashed line, if present. 
Possibility (iii) is ruled out for the diagrams to $\Sigma^{cc}$\,,
therefore these contain at least one
{\em closed path} which contributes a factor $\kappa$ to the self
energy's order of magnitude. Diagrams showing exactly one such path are easily
found (see Fig.\ \ref{fig-kondself} bottom), and it follows that in
general $\Sigma^{cc}\sim(\kappa)^1$\,. In quite the same way it is seen
that $\Sigma^{cf}\sim(\kappa)^1$ and $\Sigma^{fc}\sim(\kappa)^1$\,. For
the latter a {\em closed path} is constructed 
starting on an incoming line of an arbitrary chosen internal vertex and
proceeding in opposite direction to arrows. The situation is altered if 
$\Sigma^{ff}$
is considered: Here a path can leave the diagram (case (iii)), that is
$\Sigma^{ff}$ may contain diagrams of order $\sim(\kappa)^0=1$\,.
Actually such diagrams do exist (e.g.\ those shown in Fig.\
\ref{fig-kondself}\,top), thus $\Sigma^{ff}\sim(\kappa)^0$\,.
%
%
%
%%%**************** Appendix starts here ***
\section{Equations for Projected Spectra}
\label{app-defect}
A set of self-consistency equations is derived for the Projected Spectra
introduced in Section \ref{sec-kondo}\,.
According to the definition of $\overline{\overline{\rho}}^f$ in
Eq.(\ref{s4-aux-f}) we consider 
\begin{eqnarray*}
  \lefteqn{\widetilde{ \rho}^f(\omega)
           [1-f(\omega)] = }
    \\ 
  & & \;\;\; 
    \left| \widetilde{ G}^f(\omega) \right|^2 
    \left( {\textstyle -\frac{1}{\pi}} \right)
    \mbox{Im}\,
    \Sigma^{ff}(\omega + i0_+)
    [1 - f(\omega)]
    \;,\;
\end{eqnarray*}
where Eq.(\ref{s4-tma-gtil}) has been used with $z=\omega+i0_+$\,. 
The self energy's imaginary part is taken from Eq.(\ref{s4-sigff2}) at
$i\omega_l\to\omega+i0_+$\,, 
\begin{eqnarray}   \label{app-def-imsig1}
  \lefteqn{ \widetilde{ \rho}^f(\omega) [1-f(\omega)] = 
    \left| \widetilde{ G}^f(\omega) \right|^2 }
    \\ 
  & &  \nonumber \;\;\; 
    \times\, V^2 \int d\varepsilon\,
    \rho^c(\varepsilon) \widetilde{ \rho}^\Pi(\varepsilon+\omega) 
    [-f(\varepsilon)-g(\varepsilon+\omega)]
    f(-\omega)
    \;,\; 
\end{eqnarray}
with $1-f(\omega)=f(-\omega)$\,. 
For further transformation of expressions like this, a couple of
relations involving Bose and Fermi functions (designated as $g$ and $f$) is
useful, 
\begin{mathletters}  \label{app-def-bosferm}
  \begin{eqnarray}
    g(x-y) & = &  \label{app-def-bose}
      f(x) [1-f(y)] / [f(y)-f(x)]
      \;,\;
      \\ 
    f(x-y) & = &  \label{app-def-fermi}
      f(x) [1+g(y)] / [g(y)+f(x)]
      \;\,. 
  \end{eqnarray}
\end{mathletters}
For use in Eq.(\ref{app-def-imsig1}) the former is applied to
$g(\varepsilon+\omega)$ in the first row of the right hand side in 
\begin{eqnarray*}
  [\ldots] f(-\omega) & = & 
    [-f(\varepsilon)-g(\varepsilon+\omega)]
    [f(-\omega)-f(\varepsilon)]
    \\
  & & \hspace*{ 4mm}
    \mbox{}+ 
    [-f(\varepsilon)-g(\varepsilon+\omega)] f(\varepsilon)
    \\
  & = & 
    f(\varepsilon)[-1-g(\varepsilon+\omega)]
    \;\,.
\end{eqnarray*}
Thereby Eq.(\ref{app-def-imsig1}) becomes 
\begin{eqnarray}   \label{app-def-imsig2}
  \lefteqn{ \widetilde{ \rho}^f(\omega) [1-f(\omega)] = 
    \left| \widetilde{ G}^f(\omega) \right|^2 }
    \\ 
  & &  \nonumber \;\;\; 
    \times\, V^2 \int d\varepsilon\,
    \rho^c(\varepsilon)
    f(\varepsilon)\,
    \widetilde{ \rho}^\Pi(\varepsilon+\omega)
    [-1-g(\varepsilon+\omega)]
    \;\,.
\end{eqnarray}
By multiplying $\widetilde{Z}/{Z}$ to both sides Equation 
(\ref{s4-def-f}) is obtained immediately. 

Its counterpart Eq.(\ref{s4-def-pi}) is derived in a similar fashion: 
With Eq.(\ref{s4-pifull}) the spectrum $\widetilde{ \rho}^\Pi(\omega)$
is written 
\begin{displaymath}
  \widetilde{ \rho}^\Pi = 
    -\case{1}{\pi} \mbox{Im}\, \left(\widetilde{ \Pi}\right) = 
    \left| \widetilde{ \Pi} \right|^2
    \left\{ \frac{ -\text{Im}\,( \widetilde{ \pi} ) / \pi }
                 { | \widetilde{ \pi} |^2 }
            - \case{1}{\pi} \mbox{Im}\, \left(\widetilde{ \sigma}\right)
         \right\}
    \;\,. 
\end{displaymath}
$\widetilde{ \pi}$ and $\widetilde{ \sigma}$ are taken directly from
the graphical definition shown in Fig.\ \ref{fig-ladders}\,,
\begin{mathletters}    \label{app-def-org}
  \begin{eqnarray}
    \widetilde{ \pi}(z) & = &   \label{app-def-piorg}
      \int\!\!\!\int \!\! d \varepsilon d \varepsilon'\,
      \widetilde{ \rho}^f(\varepsilon)
      \widetilde{ \rho}^f(\varepsilon')
      \frac{ f(\varepsilon) - f(-\varepsilon') }
           { z-\varepsilon-\varepsilon' }
      \;,\;
      \\ \nonumber \\ 
    \widetilde{ \sigma}(z) & = &   \label{app-def-siorg}
      N_J V^2 \!\!
      \int\!\!\!\int \!\! d \varepsilon d \varepsilon'\,
      \rho^c(\varepsilon)
      \widetilde{ \rho}^f(\varepsilon')
      \frac{ f(\varepsilon) - f(-\varepsilon') }
           { z-\varepsilon-\varepsilon' }
      \;\,. 
  \end{eqnarray}
\end{mathletters}
The effect
of a prefactor $[-1-g(\omega)]$ to Eqs.(\ref{app-def-piorg}) and
(\ref{app-def-siorg}) is made explicit by application of 
Eqs.(\ref{app-def-bosferm})\,.
Collecting the resulting parts together, we get 
\begin{eqnarray*}
  \lefteqn{ \widetilde{ \rho}^\Pi(\omega)
            [-1-g(\omega)] = 
            \left| \widetilde{ \Pi}(\omega) \right|^2 }
    \\ 
  & & \;\;\; 
    \times \int d\varepsilon\,
    \left\{ \frac{ \widetilde{ \rho}^f(\varepsilon)[1-f(\varepsilon)] }
                 { \left| \widetilde{ \pi}(\omega) \right|^2 }
            + N_J V^2 \rho^c(\varepsilon)[1-f(\varepsilon)]
            \right\} 
    \\ 
  & & \;\;\;  \hspace*{ 14mm}
    \times\, \widetilde{ \rho}^f(\omega-\varepsilon)
           [1-f(\omega-\varepsilon)]
    \;\,. 
\end{eqnarray*}
Since $|\widetilde{ \pi}|^2\gg\kappa$ the first term under the braces may
be neglected with Eq.(\ref{s4-rhosmall})\,, and Eq.(\ref{s4-def-pi})
appears in writing  
$\widetilde{Z}/Z$ on both sides of the above formula. Furthermore, the 
result confirms the order of magnitude Eq.(\ref{s4-pismall}) directly for
the special case of the ladder approximation. 
%
%
%
%%%**************** Appendix starts here ***
\section{Some Exact Relations}
\label{app-sumrules}
%
%
%%%**********************************
\subsection{Sum Rules}
\label{app-sumrules-sumrules}
Exact sum rules are proven for spectra $\widetilde{ \rho}^f$\,,    
$\widetilde{\rho}^\Pi$\,, 
$\overline{\overline{\rho}}^f$\,, $\overline{\overline{\rho}}^\Pi$\,,
and the physical $f$-spectrum $\rho^f$\,. These are  verified
within the Self-Consistent T-Approximation. Results have already
been stated in Section \ref{sec-kondo}  through
Eqs.(\ref{s4-sum-all})\,. 
The spin $\sigma=\pm 1$ is  arbitrary in the following, since no external
magnetic field is considered. 

The total spectral weight of the propagator $\widetilde{ G}^f$ is 
derived from Eq.(\ref{s3-gtilff}) as 
$\langle \{ f_\sigma, f^\dagger_\sigma \} \rangle = 1$\,, 
accordingly the corresponding spectrum defined in Eq.(\ref{s4-rhotildef})
fulfills the sum rule Eq.(\ref{s4-sum-ftil})\,. 
In the same fashion, the spectral weight of $\widetilde{ \Pi}$ follows with 
Eqs.(\ref{s4-pidef}), (\ref{s4-rhopi}) in the form 
\begin{eqnarray*}
  \lefteqn{ \int d\varepsilon\,
            \widetilde{ \rho}^\Pi(\varepsilon) = 
            \widetilde{ \Pi}(0_-) - \widetilde{ \Pi}(0_+)   }
    \\
  \;\;\; & = & 
    \langle n^f_\uparrow n^f_\downarrow \widetilde{ \rangle} - 
    \langle (1-n^f_\uparrow)(1-n^f_\downarrow) \widetilde{ \rangle}
    = 
    \sum_\sigma \langle n^f_\sigma \widetilde{ \rangle} - 1
    \;,\;
\end{eqnarray*}
and by introducing $\widetilde{ \rho}^f$ via
$  \langle n^f_\sigma \widetilde{ \rangle} = 
    \int d \varepsilon\,
    \widetilde{ \rho}^f(\varepsilon) f(\varepsilon)$\,, 
the sum rule Eq.(\ref{s4-sum-pi}) is confirmed. 

In order to derive a sum rule for the Projected Spectra we first
note the physical $f$-spectral-weight factor per spin, 
\begin{eqnarray}  \label{app-sum-fnorm1}
  \langle 1-n^f_\sigma \rangle & = &
    - (\widetilde{ Z}/Z) \widetilde{ G}_\sigma^f(0_+) 
    \\ 
  & = &  \nonumber 
    \frac{\widetilde{ Z}}{Z} \int d \varepsilon\,
    \widetilde{ \rho}^f(\varepsilon)
    [1 - f(\varepsilon)] 
    = 
    \int d \varepsilon\,
    \overline{\overline{\rho}}^f(\varepsilon)
    \;\,. 
\end{eqnarray}
Consider also the probability of the local $f$-orbital being empty,
\begin{eqnarray}    \label{app-sum-empty}
  \lefteqn{ \langle (1-n^f_\uparrow)(1-n^f_\downarrow) \rangle
            = (\widetilde{ Z}/Z) \widetilde{ \Pi}(0_+)
            } 
    \\ 
  & \;\;\; = &  \nonumber
    \frac{\widetilde{ Z}}{Z} \int d \varepsilon\,
    \widetilde{ \rho}^\Pi(\varepsilon)
    [-1-g(\varepsilon)]
    = 
    \int d \varepsilon\,
    \overline{\overline{\rho}}^\Pi(\varepsilon)
    \;\,. 
\end{eqnarray}
In the last two expressions the respective first equality sign
stems from the fact that
states with doubly occupied $f$-orbital do not contribute to the traces
(this has been explained in Section \ref{sec-wick}\,). 
At $U\to\infty$ the probability of double $f$-occupancy is zero, 
and with the probability
$\displaystyle \langle n^f_\sigma(1-n^f_{-\sigma}) \rangle$
for single $f$-occupancy a completeness relations holds,
\begin{displaymath}
  \langle (1-n^f_\uparrow)(1-n^f_\downarrow) \rangle + 
    \sum_\sigma\langle n^f_\sigma(1-n^f_{-\sigma}) \rangle = 1
    \;\,. 
\end{displaymath}
This leads with Eqs.(\ref{app-sum-fnorm1}) and (\ref{app-sum-empty}) 
to the sum rule Eq.(\ref{s4-sum-defect}) stated in Section
\ref{sec-kondo}\,. 

The physical $f$-excitation spectrum is normalized according to
Eq.(\ref{s2-fgf2})\,, i.e.\ 
\begin{displaymath}
  \int d \varepsilon\,
  \rho^f(\varepsilon) = 
    F_\sigma(0_-) - F_\sigma(0_+) =
    \langle 1-n^f_{-\sigma} \rangle 
    \;\,. 
\end{displaymath}
By use of Eq.(\ref{app-sum-fnorm1})\,, the sum rule
Eq.(\ref{s4-sum-fnorm}) is evident. 

A verification of sum rules in SCTA
is performed for $\widetilde{ \rho}^f$ and $\widetilde{ \rho}^\Pi$
via the behavior of corresponding propagators at infinity, 
\begin{displaymath}
  |z|\to\infty\,: \hspace*{ 2mm}
    \widetilde{ G}^f(z) \to a^f / z
    \;,\;
    \widetilde{ \Pi}(z) \to a^\Pi / z
    \;\,. 
\end{displaymath}
From the self-consistency equations (\ref{s4-tma-all}) at
$|z|\to\infty$ we get  
\begin{displaymath}
  a^f=1
    \;,\;
    a^\Pi = \int d \varepsilon\,
            \widetilde{ \rho}^f(\varepsilon)
            [2f(\varepsilon)-1]
            \;,\;
\end{displaymath}
i.e.\ the values from exact sum rules Eqs.(\ref{s4-sum-ftil}) and
(\ref{s4-sum-pi}) are reproduced.  
The $f$-spectrum's normalization in SCTA is obtained via 
integration of Eq.(\ref{s4-rhof})\,, 
\begin{eqnarray*}
  \int d \omega\, \rho^f(\omega) & = & 
    \int d \omega\, \overline{\overline{\rho}}^f(\omega)\,
                    \int d \varepsilon 
                    \widetilde{ \rho}^\Pi(\varepsilon) + 
    \int d\omega\,  \overline{\overline{\rho}}^\Pi(\omega)
    \\ 
  & = &
    3 \int d \omega\,
    \overline{\overline{\rho}}^f(\omega) - 1
    \;\,. 
\end{eqnarray*}
Corrections $\sim(\kappa)^1$ from application of the sum rule
Eq.(\ref{s4-sum-pi}) are negligible in the Kondo regime. Obviously
the norm of $\rho^f$ does not match the sum rule
Eq.(\ref{s4-sum-fnorm})\,. 
%
%
%%%**********************************
\subsection{Spectral Decomposition at Zero Temperature}
\label{app-sumrules-spectra}
Consider the unphysical spectra
$\widetilde{\rho}^f$ and $\widetilde{\rho}^\Pi$ at zero temperature
$T=0$\,. A spectral decomposition of $\widetilde{ G}^f_\sigma$ leads
to 
$ \widetilde{\rho}^f_\sigma(\varepsilon) = 
    \widetilde{\rho}^{f(+)}_\sigma(\varepsilon) + 
    \widetilde{\rho}^{f(-)}_\sigma(\varepsilon) $
where
\begin{displaymath}
  \widetilde{\rho}^{f(\pm)}_\sigma(\varepsilon) = 
    \sum _\alpha \left| \langle \Psi_\alpha | f^{(\dagger)}_\sigma |
                        \Psi_0 \rangle \right|^2
    \delta( \varepsilon\mp [E_\alpha-E_0])
\end{displaymath}
at $T\to 0$\,. An $f$-creation operator $f^\dagger_\sigma$ is used in
$\widetilde{\rho}^{f(+)}_\sigma$\,. 
The ground state $|\Psi_0\rangle$ of the Hamiltonian
Eq.(\ref{s2-htil}) in the full Fock space of $f$- and conduction
electrons is 
$|\Psi_0\rangle = 
   |2;\mbox{FS}^c\rangle$\,,
a doubly occupied $f$-orbital plus the filled Fermi sea of conduction
electrons. It is unaffected by the perturbation in
Eq.(\ref{s2-htil})\,, and its energy $E_0=2 \varepsilon^f + E^c$ is
significantly lower than the ground state energy $E_G\approx
E^c+\varepsilon^f - T_K$ of the 
`physical' subspace ${\cal H}_{0,1}$ with $f$-occupancy $\le 1$\,; see
also Sect.\ \ref{sec-small}\,. Since 
$f^\dagger_\sigma|2;\mbox{FS}^c\rangle\equiv 0$
and
$f_\sigma|2;\mbox{FS}^c\rangle = 
   \pm |-\sigma;\mbox{FS}^c\rangle$\,,
it follows $\widetilde{\rho}^{f(+)}_\sigma(\varepsilon)=0$ and
\begin{displaymath}  
  \widetilde{\rho}^f_\sigma(\varepsilon) = 
    \sum_i
    \left| \langle\Psi_i|-\sigma;\mbox{FS}^c\rangle\right|^2
    \delta(\varepsilon-[2\varepsilon^f+E^c-E_i])
    \;\,. 
\end{displaymath}
From the complete set $\{|\Psi_\alpha\rangle\}$ of eigenstates only
the subspace $\{|\Psi_i\rangle\}={\cal H}_{0,1}$ contributes to the
overlap integral, with
energies $E_i\ge E_G$\,. Accordingly the spectrum
$\widetilde{\rho}^f_\sigma(\varepsilon) \sim
   \Theta(E^{Th}-\varepsilon)$
shows a threshold at energy $E^{Th}=2 \varepsilon^f-E_G+E^c$
slightly above the bare $f$-level $\varepsilon^f$\,. The same
threshold emerges in
$\widetilde{\rho}^\Pi(\varepsilon) \sim
   \Theta(E^{Th}-\varepsilon)$\,: 
From a similar analysis $\widetilde{\rho}^\Pi(\varepsilon)$ is obtained
in the form of $\widetilde{\rho}^f$ with 
$|-\sigma;\mbox{FS}^c\rangle$
replaced by
$|0;\mbox{FS}^c\rangle$\,. 
{\em At zero temperature}\/ the unphysical spectra
$\widetilde{\rho}^f_\sigma$ and $\widetilde{\rho}^\Pi$ 
match the so-called `ionic spectra' $\rho_\sigma$ and $\rho_0$ of 
Resolvent-Perturbation Theory \cite{note-bic} via
\begin{equation}  \label{app-sum-match}
  \rho_\sigma(\varepsilon) = 
    \widetilde{\rho}^f_\sigma(2 \varepsilon^f-\varepsilon)
    \;,\;
  \rho_0(\varepsilon) = 
    \widetilde{\rho}^\Pi(2 \varepsilon^f-\varepsilon)
    \;\,.
\end{equation}
Accordingly an algebraic singularity is expected in
$\widetilde{\rho}^{f/\Pi}$ at energy approaching $E^{Th}$ from below. 
It has nevertheless to be noted that the strict equality
Eq.(\ref{app-sum-match}) is lost at any finite temperature and is likely
to be lost if approximations are considered (e.g.\ SCTA vs.\ NCA). Also is a
relation like Eq.(\ref{app-sum-match}) {\em not} found for the
Projected Spectra introduced in Eq.(\ref{s4-aux-all}) and the
so-called Defect Spectra \cite{mue84} (also called e.g.\
`negative-frequency spectra'\cite{biccoxwil87}) of NCA theory. 
%
%
%
%%%********************* End of body of paper ***************************
%

%
%%%****************** figures follow here ********************************
%

%
\newlength{\mysize} 
\def\loadepsfig#1{
 \def\figname{#1}
 \vbox to 10pt {\ }
 \vbox{ \hbox to \hsize {
   \mysize\hsize    \advance \mysize by -20pt 
   \def\epsfsize##1##2{\ifdim##1>\mysize\mysize\else##1\fi}
   \hfill \epsffile{\figname.eps} \hfill 
        } }
 \vbox to 7pt {\ }
 }
%
%
%%%******************** to be used in figure-tex-sources:
%%%*** gloscale is scale-factor for xfig-figures 
%

%  In galley-style, we need eps-figures:
\def\mydot{\rule{0.25pt}{0.25pt}}
%  macro fignum is dummy here:
\def\fignum#1{ \mbox{} }
\def\gloscale{1.0}
\def\dispfig#1{
  \def\figname{#1}
  \def\epsfsize##1##2{\gloscale##1}
  \settowidth{\mysize}{ \epsffile{\figname.eps} }
  \parbox{\mysize}{ \epsffile{\figname.eps} }
  }
%
%
%%%*************** End of  figdefs.tex ***********************************
%%%***********************************************************************
%

%

%
%
%%%************** Figure vertgf :
%
\begin{figure}
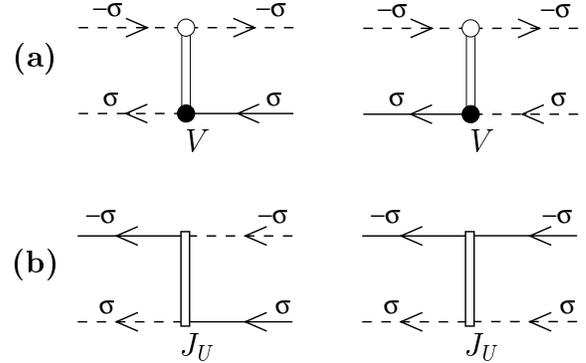
 

 \loadepsfig{vertgf}
%%%*** figure built from vertgf.eps from vertgf.tex, see there

 %
 \caption[\ ]{ 
Vertices of the Feynman-diagram expansion. Dashed lines correspond to
local Green's functions \protect$\widetilde{ G}^{f(0)}\protect$\,,
Eq.(\protect\ref{s2-gtil0}), 
full lines stand for conduction-electron propagators
\protect$G^c\protect$\,, 
Eq.(\ref{s2-gc0})\,. Vertices {\bf (a)} originate from
\protect$H^V_{01}\protect$ (see 
Eq.(\protect\ref{s2-vpart01})) and are interpreted
in the subspace without doubly occupied local
\protect$f\protect$-orbital: A transfer band- to
\protect$f\protect$-electron (left) or vice versa (right) is permitted
only if a \protect$f\protect$-hole of opposite spin is present. In
{\bf (b)} the spin-flip and charge scattering vertex of
\protect$H^J\protect$ (see Eq.(\protect\ref{s2-exham})) are shown on
the left and right, respectively. 
         }
 \label{fig-vertgf}
\end{figure}

%

%
%
%%%************** Figure logloops :
%
\begin{figure}

 \loadepsfig{logloops}
%%%*** figure built from logloops.eps from logloops.tex, see there

 %
 \caption[\ ]{
Simple ladder diagrams from particle--particle \protect$(pp)\protect$
and particle--hole \protect$(ph,\overline{ph})\protect$ vertex
contributions to the \protect$f\protect$-Green's function at
\protect$U\to\infty\protect$\,. These may show
logarithmic divergences, see text. 
         }
 \label{fig-logloops}
\end{figure}

%

%
%%%************** Figure matrix :
%
\begin{figure} 

 \loadepsfig{matrix}
%%%*** figure built from matrix.eps from matrix.tex, see there

 %
 \caption[\ ]{ 
{\bf (a)}:\,~Notation of renormalized  Green's functions occurring
in skeleton diagrams of 
the \protect$U\to\infty\protect$ problem. A definition is given in
Eq.(\protect\ref{s3-allgtil})\,. {\bf
(b)}:\,~\protect$2\times 2\protect$-matrix-Green's function (left),
its matrix elements are shown in {\bf (a)}\,, and corresponding
bare propagator 
(right)\,. {\bf (c)}:\,~Dyson's equation involving matrix-self energy
Eq.(\protect\ref{s3-matrix-se})\,. } 
 \label{fig-matrix}
\end{figure}

%

%
%
%%%************** Figure matvert :
%
\begin{figure}

 \loadepsfig{matvert}
%%%*** figure built from matvert.eps from matvert.tex, see there

 %
 \caption[\ ]{
{\bf (a)}:\,~Vertex used with matrix-Green's function from Fig.\
\protect\ref{fig-matrix}~(b) for \protect$U\to\infty\protect$\,. It
collects both hybridization types shown in Fig.\
\protect\ref{fig-vertgf}~(a)\,. `Spinor' indices
\protect$a^{(')}, b^{(')}=1^{(')},2^{(')}\protect$ indicate
canonical operators \protect$1^{(')}\equiv f^{(\dagger)}\protect$\,,
\protect$2^{(')}\equiv C^{(\dagger)}\protect$\,. 
{\bf (b)}:\,~Vertex combination used in
expansion of matrix-self energy
\mbox{\boldmath\protect$\Sigma\protect$}\,. 
         }
 \label{fig-matvert}
\end{figure}

%

%
%
%%%************** Figure selfbethe :
%
\begin{figure}

 \loadepsfig{selfbethe}
%%%*** figure built from selfbethe.eps from selfbethe.tex, see there

 %
 \caption[\ ]{
Exact skeleton representation of matrix-self energy at
\protect$U\to\infty\protect$\,, with
full Green's function and vertex from Figs.\ \protect\ref{fig-matrix}~(b) and
\protect\ref{fig-matvert}~(b)\,. Due to the vertex' spin
restriction no `Fock diagram' is present. All higher orders are
collectively expressed through the irreducible vertex function
\protect$\Gamma^2_P\protect$\,. 
         }
 \label{fig-selfbethe}
\end{figure}

%

%
%
%%%************** Figure selftma :
%
\begin{figure} 

 \loadepsfig{selftma}
%%%*** figure built from selftma.eps from selftma.tex, see there

 %
 \caption[\ ]{
Self energy in Self-Consistent T-Approximation. With spin
restriction at vertices, all exchange parts present in the vertex function
in Fig.\ \protect\ref{fig-selfbethe} do not contribute to
\protect$\bbox{\Sigma}\protect$ and internal spin sums do not
occur. The self energy is exact to \protect$V^2\protect$ (first
row), higher orders separate into ladder sums from 
\protect$(pp)\protect$-channel (2nd row) and \protect$(ph)\protect$-
(3rd row) and \protect$(\overline{ph})\protect$-channels (4th row). 
         }
 \label{fig-selftma}
\end{figure}

%

%
%
%%%************** Figure omitted :
%
\begin{figure}

 \loadepsfig{omitted}
%%%*** figure built from ... , see there

 %
 \caption[\ ]{
Examples of diagrams to \protect$\Sigma^{ff}\protect$ which do not
contribute in the Kondo regime: 
\ {\bf (a)}:\,~due to the {\em loop theorem} (possible closed paths are
indicated by dotted lines);
\ {\bf (b)}:\,~due to the presence of \protect$\widetilde{
G}^{fc}\protect$ or \protect$\widetilde{ G}^{cf}\protect$\,. 
Notation follows Fig.\ \protect\ref{fig-matrix}~(a)\,. 
         }
 \label{fig-omitted}
\end{figure}

%

%
%
%%%************** Figure kondself :
%
\begin{figure}
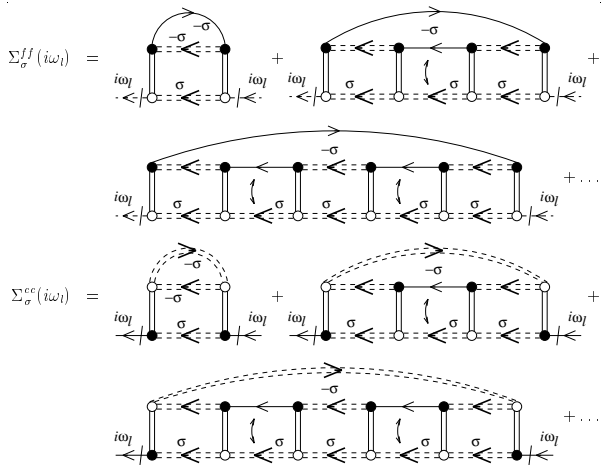


 \loadepsfig{kondself}
%%%*** figure built from ... , see there

 %
 \caption[\ ]{
Self energy
\protect$\bbox{\Sigma}\protect$ in the Kondo regime, consisting only
of \protect$(pp)\protect$-type diagrams with Green's functions
\protect$\widetilde{ G}^f\protect$ (dashed double line) and bare
conduction electrons \protect$G^c\protect$ (full lines). Off-diagonal
elements 
\protect$\Sigma^{cf}\protect$ and \protect$\Sigma^{fc}\protect$ as
well as all \protect$(ph)\protect$- and
\protect$(\overline{ph})\protect$-type diagrams included in 
Fig.\ \protect\ref{fig-selftma} are negligible
in the Kondo regime, examples are shown in Fig.\
\protect\ref{fig-omitted}\,.
In each plaquette indicated by a small double arrow, parallel band-
and \protect$f\protect$-lines may be interchanged, leading to the full
set of diagrams to be considered. 
         }
 \label{fig-kondself}
\end{figure}

%

%
%
%%%************** Figure ladders :
%
\begin{figure}
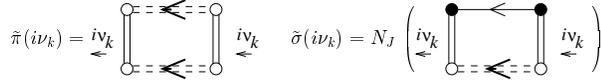


 \loadepsfig{ladders}
%%%*** figure built from ... , see there

 %
 \caption[\ ]{
Definition of `ladder elements' occurring in Fig.\
\protect\ref{fig-kondself}\,. Spin degeneracy leads to the prefactor
\protect$N_J\equiv 2\protect$ in 
\protect$\widetilde{ \sigma}(i\nu_k)\protect$\,. 
         }
 \label{fig-ladders}
\end{figure}

%

%
%
%%%************** Figure spectra :
%
\begin{figure}
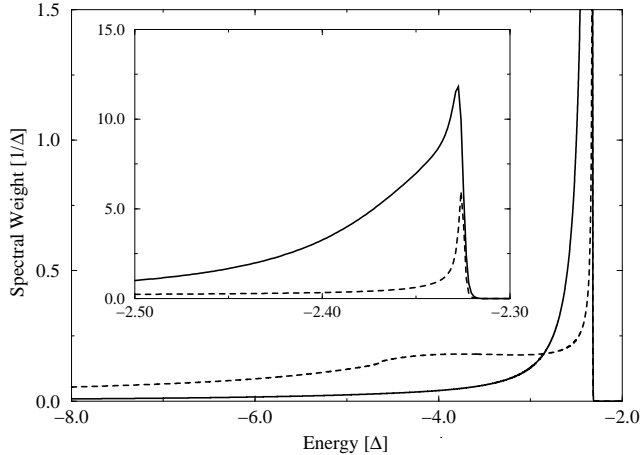


 \loadepsfig{spectra}
%%%*** figure built directly by xmgr (Parameter file: spectra*.xmgr)

 %
 \caption[\ ]{
Spectra \protect$\widetilde{\rho}^f\protect$\, (full line) and
\protect$\widetilde{\rho}^\Pi\protect$\, (dashed line) from numerical 
solution of SCTA equations (see text). Parameters are 
\protect$\varepsilon^f=-3.0\Delta\protect$\,,
\protect$D=10.0\Delta\protect$\,, \protect$N_J\equiv 2\protect$\,, and
\protect$k_B T=0.1T_K\protect$\,. The spectral weight turns to zero for
energies above the threshold. 
{\bf Inset:} Neighborhood of the threshold 
(note the different vertical scale). 
         }
 \label{fig-spectra}
\end{figure}

%

%
%
%%%************** Figure fdos :
%
\begin{figure}
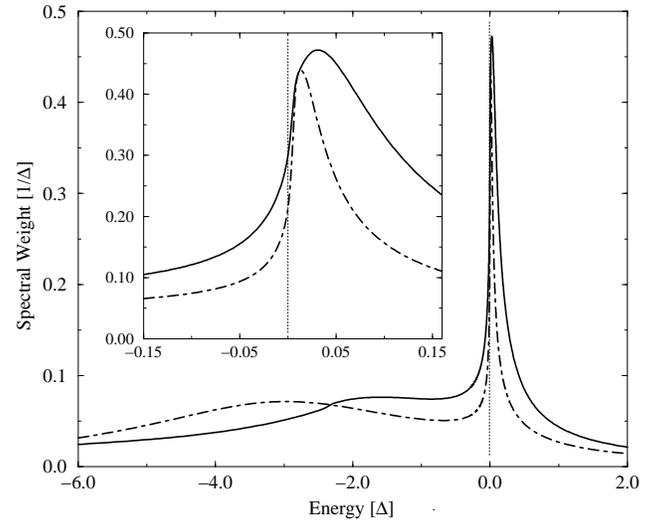


 \loadepsfig{fdos}
%%%*** figure built directly by xmgr (Parameter file: fdos*.xmgr)

 %
 \caption[\ ]{
Local \protect$f\protect$-excitation spectrum of the
\protect$U\to\infty\protect$ Anderson model from Self-Consistent 
T-Approximation (full line) and NCA (dashed dotted line). 
{\bf Inset:} Abrikosov--Suhl resonance near the Fermi energy $\omega= 0$\,. 
Parameters as in Fig.\ \protect\ref{fig-spectra}\,. 
         }
 \label{fig-fdos}
\end{figure}

%
%%%************************** End of  Figures ***************************
%
%

\begin{thebibliography}{100}
\bibitem{pap0995-note1}
Strictly speaking, we have to use an unusual convergence factor
  $\exp(-i\omega_l 0_+)$ for a loop built from a single dashed line at a vertex
  in Fig.\ \ref{fig-vertgf} (a)\,, and a factor $\exp(i\omega_l 0_+)$ for a
  single dashed or full line closed onto itself at a vertex Fig.\
  \ref{fig-vertgf} (b) right. Other kinds of simple loops do not occur. In
  order to get the `global' rule using only $\exp(-i\omega_l 0_+)$ as stated in
  the text, we have to turn operators in the second term of $H^J$\,,
  Eq.(\ref{s2-exham})\,, to non-normal ordering $f f^\dagger$ and $c c^\dagger$
  before applying Wick's theorem. This results in a potential scattering for
  band electrons and a small shift of $f$-level $\varepsilon^f$ which may be
  ignored.
\bibitem{pap0995-note5}
The discussion in Note \onlinecite{pap0995-note1} remains valid for skeleton
  diagrams, since the only closed fermion loop to be considered in addition is
  that built of a single mix-Green's function $\widetilde{G}^{fc/cf}_\sigma$\,:
  It's spectral function is normalized to Null, therefore a convergence factor
  has no effect and is arbitrary.
\bibitem{pap0995-note7}
For the analytical solution of mean-field equations at $T\to 0$ and
  $V\ll|\varepsilon^f|<D$\,, $\varepsilon^f<0$ one may rely on slave-boson
  treatments (see e.g.\ Ref.\ \onlinecite{newrea87}\,). Our mean-field
  equations can be reproduced formally from the slave-boson equations
  \cite{newrea87}\,, if the degeneracy $N_J$ in the latter is set to $N_J=1$\,.
  Thus a trivial solution is to be expected in the present case.
\bibitem{pap0995-note8}
This is a consequence of $1+g(\omega)\sim\mbox{sgn}(\omega)$ and
  $\widetilde{\rho}^\Pi(\omega)\sim-\mbox{sgn}(\omega)$\,. The latter follows
  from the fact that $\widetilde{\Pi}$ defined in Eq.(\ref{s4-pidef}) is a
  bosonic double-time Green's function of the form $+\langle{\cal
  T}\{A(\tau)A^\dagger(\tau')\} \rangle$\,.
\end{thebibliography}
\end{document}